\RequirePackage{ifpdf}
\documentclass[11pt]{article}

\usepackage{jheppub}
\usepackage{multicol}
\usepackage{graphicx}
\usepackage{amsmath}
\usepackage{url}
\usepackage{multirow}
\usepackage{rotating}
\usepackage{afterpage}
\usepackage{lmodern}
\usepackage[T1]{fontenc}

\usepackage{pdflscape}
\usepackage[table]{xcolor}
\usepackage{caption}
\usepackage{subcaption}
\usepackage{aas_macros}
\usepackage{jdefs}
\usepackage{bm}
\usepackage{ragged2e}
\usepackage{changepage}
\usepackage{booktabs}
\usepackage{cleveref}
\usepackage{xspace}
\usepackage{bm}

\definecolor{row-highlight}{rgb}{0.97, 0.51, 0.47}
\newcommand{\code}{\texttt}

\newcommand{\neff}{\ensuremath{n_\text{eff}}}
\newcommand{\nlike}{\ensuremath{n_\mathcal{L}}}
\renewcommand{\vec}{\bm}
\newcommand{\given}{\,|\,}

\newcommand{\rhat}{\ensuremath{\hat R}\xspace}

\newcommand{\nlive}{\ensuremath{n_\text{live}}}

\newcommand{\adelaide}{ARC Centre for Dark Matter Particle Physics, Department of Physics, University of Adelaide, Adelaide, SA 5005, Australia}

\newcommand{\xjtlu}{Department of Physics, School of Mathematics and Physics, Xi'an Jiaotong-Liverpool University, Suzhou, 215123, China}

\newcommand{\monash}{School of Physics and Astronomy, Monash University, Melbourne, VIC 3800, Australia}

\newcommand{\cambridge}{Cavendish Laboratory, University of Cambridge, JJ Thomson Avenue, Cambridge, CB3 0HE, UK}
\newcommand{\kicc}{Kavli Institute for Cosmology, Madingley Road, Cambridge, CB3 0HA, UK}

\newcommand{\ific}{Instituto de F\'isica Corpuscular, IFIC-UV/CSIC, Carrer del Catedrátic José Beltrán Martinez, 2, Valencia, Spain}

\newcommand{\caltech}{Cahill Center for Astronomy and Astrophysics, California Institute of Technology, Pasadena, CA 91125, USA}
\newcommand{\leiden}{Leiden Observatory, Leiden University, PO Box 9513, 2300 Leiden, The Netherlands}

 \newlength{\wth}
\title{A comparison of Bayesian sampling algorithms for high-dimensional particle physics and cosmology applications}

\abstract{For several decades now, Bayesian inference techniques have been applied to theories of particle physics, cosmology and astrophysics to obtain the probability density functions of their free parameters. In this study, we review and compare a wide range of Markov Chain Monte Carlo (MCMC) and nested sampling techniques to determine their relative efficacy on functions that resemble those encountered most frequently in the particle astrophysics literature. Our first series of tests explores a series of high-dimensional analytic test functions that exemplify particular challenges, for example highly multimodal posteriors or posteriors with curving degeneracies. We then investigate two real physics examples, the first being a global fit of the $\Lambda$CDM model using cosmic microwave background data from the Planck experiment, and the second being a global fit of the Minimal Supersymmetric Standard Model using a wide variety of collider and astrophysics data. We show that several examples widely thought to be most easily solved using nested sampling approaches can in fact be more efficiently solved using modern MCMC algorithms, but the details of the implementation matter. Furthermore, we also provide a series of useful insights for practitioners of particle astrophysics and cosmology.}

\collaboration{The DarkMachines High Dimensional Sampling Group}  % jhep style
\author[1, 2]{Joshua Albert}
\author[3]{Csaba~Bal\'azs,}
\author[4]{Andrew Fowlie,}
\author[5,6]{Will Handley,}
\author[7]{Nicholas Hunt-Smith,}
\author[8]{Roberto Ruiz de Austri,}
\author[7]{Martin~White}  % changed to jhep style

\affiliation[1]{\caltech}
\affiliation[2]{\leiden}
\affiliation[3]{\monash}
\affiliation[4]{\xjtlu}
\affiliation[5]{\cambridge}
\affiliation[6]{\kicc}
\affiliation[7]{\adelaide}
\affiliation[8]{\ific}

\emailAdd{martin.white@adelaide.edu.au}
\emailAdd{nicholas.hunt-smith@adelaide.edu.au}

\begin{document}

\maketitle


\section{Introduction}

Global fitting is an essential tool across physics, allowing us to combine all relevant experimental data and learn the unknown parameters of a physics model~\cite{AbdusSalam:2020rdj,gambit,hepfit,MasterCodemSUGRA,NuFIT}.
The range of applications includes cosmology~\cite{2110.11421, 2002.11406}, astrophysics~\cite{2002.11406, Chang:2022jgo, Chang:2023cki, GAMBIT:2021rlp}, and nuclear and particle physics.  In the latter field global fits are used in quantum chromodynamics~\cite{2302.12124, 2206.12465, 2201.06586, 2106.06390}, neutrino physics~\cite{2111.03086, 2007.14792, 2007.08526, 2003.08511}, effective field theories~\cite{2207.10088, 2103.14022, 2012.02779, 2003.05432}, and in beyond the Standard Model phenomenology~\cite{2210.16337, 2207.05103, 2202.11753, 2101.07811, 2012.02799, 2007.02985}.  

There are two dominant approaches to global fitting that are widely used across all sciences: frequentist and Bayesian inference. The latter is particularly prevalent in particle astrophysics, astrophysics and cosmology, and common in particle physics.  
In particle physics, Bayesian inference is widely used for example in non-perturbative quantum chromodynamics studies, such as fits of parton distribution functions and fragmentation functions~\cite{2201.07240, DelDebbio:2021whr, HERAFitterdevelopersTeam:2015cre}, extraction of signs of physics beyond the Standard Model~\cite{2112.07274, 2104.07680, 2101.00428, Anisha:2020ggj, 2008.01113}, analysis of effective field theories~\cite{Anisha:2020ggj, Kowalska_2019, hepfit}, and the determination of the parameters of the neutrino sector~\cite{2006.11237, DellOro:2019pqi, NuFit15}.
In particle astrophysics, applications range from neutrino astronomy~\cite{Bose:2012qh, Romero-Wolf:2017xqe, Capel:2020txc}, through to cosmic ray studies~\cite{Fowlie:2017fya}, dark matter searches and phenomenology~\cite{SSDM, SSDM2, Athron:2018hpc, Liem:2016xpm}, gravitational wave astronomy~\cite{Ashton:2021anp, BilbyMCMCGuide, Ashton:2018jfp}, and global fits and model comparison~\cite{Athron:2017ard,hepfit,Buchmueller:2007zk, NuFit15}.
In astrophysics and cosmology, Bayesian tools are used in the extraction of cosmological or particle astrophysics parameters and gravitational wave studies~\cite{2111.13083, Hergt:2021qlh, 2101.11882, 2001.05393, 2002.04035}.

In a previous work, we surveyed algorithms and codes for optimisation, which is more relevant for frequentist inference~\cite{DarkMachinesHighDimensionalSamplingGroup:2021wkt}. We compared a variety of different techniques on analytic test functions and real physics examples to determine which approaches were the best starting point for problems in particle physics, astrophysics and cosmology. The purpose of this work is to perform an analogous study for Bayesian inference techniques which, as a rule, are the most popular choice in cosmology whilst also having a wide utility in particle physics global fits. There is a significant literature of sampling algorithms for Bayesian inference, with new algorithms being provided at a steady rate. For users in the field, it is not easy to determine which technique to use for a given problem, despite the choice having major implications for the accuracy of the final results and the efficiency with which they are obtained. In the following review, we therefore compare and contrast a wide selection of Markov chain Monte Carlo and nested sampling techniques on analytic test functions and real physics examples, in order to see which approaches work best for which problems. This allows users with physical insight on the expected shape of their posterior to select the most promising technique. Throughout, we aim to compare techniques from the position of the typical user --- we do not therefore perform extensive optimisation of the hyperparameters of different algorithms, but instead stick close to default settings in order to compare their relative efficacy under reasonable test conditions.

In the Bayesian statistical framework (see e.g.,~\cite{Jeffreys:1939xee,d2003bayesian,sivia2006data} for an introduction), we begin with an observed set of data $\vec{D}$ that we wish to describe using a model that depends on some unknown parameters $\vec{\theta}$. Bayesian inference proceeds by updating a chosen prior distribution of the parameters $p(\vec{\theta})$ to a posterior distribution given the data $p(\vec{\theta}\given\vec{D})$, through application of Bayes' theorem:
\begin{equation}\label{eq:bayes}
  p(\vec{\theta}\given\vec{D})=\frac{\mathcal{L}(\vec{D}\given\vec{\theta}) \, p(\vec{\theta})}{p(\vec{D})}.
\end{equation}
Here $\mathcal{L}(\vec{D}\given\vec{\theta})$ is the likelihood function (the probability of the data given the parameter point $\vec{\theta}$) and $p(\vec{D})$ is the Bayesian evidence, which ensures that the left-hand side is normalised, as expected for a probability density function. 

In many cases, including those explored in this paper, the unnormalised posterior (or equivalently the joint distribution) $\mathcal{L}(\vec{D}\given\vec{\theta}) \, p(\vec{\theta})$ is available analytically and tractable (though there are situations in which this may not be the case~\cite{Andrieu_2009}) but the evidence $p(\vec{D})$ is not readily available. We wish to learn about one or more of the parameters through the posterior or through moments and quantiles of parameters with respect to the posterior. This becomes computationally challenging as the number of parameters grows and for multi-modal posteriors, that is, cases in which there the posterior contains more than one mode. The challenge lies in integrating the posterior with respect to model parameters; e.g., the marginal posterior,
\begin{equation}\label{eq:marginal}
    p(\theta_1 \given \vec{D}) = \int p(\vec{\theta} \given \vec{D}) \, \text{d}\theta_2 \text{d}\theta_3 \ldots \text{d}\theta_n,
\end{equation}
or the expectation,
\begin{equation}\label{eq:expectation}
    \langle g \rangle = \int g(\vec \theta) \, p(\vec{\theta} \given \vec{D}) \, \text{d}\theta_1 \text{d}\theta_2 \text{d}\theta_3 \ldots \text{d}\theta_n,
\end{equation}
despite the unavailability of the evidence in \cref{eq:bayes}.
This has spurred the creation of numerous algorithms aimed at rendering these computations more feasible.

Our paper is structured as follows. In \cref{sec:algorithms}, we provide a description of each of the algorithms we use in our study. In \cref{sec:methodology} we explain the metrics that we use to compare them, and in \cref{sec:results} we describe the examples that we use for our comparisons and detail the results.  Finally, we present conclusions in \cref{sec:conclusions}.
The appendices include supplementary materials related to the metrics we have used, as well as posterior means and standard deviations for the test functions.

\section{Sampling algorithms}\label{sec:algorithms}

Due to their current dominance in particle astrophysics and cosmology applications, we consider variants of two classes of Monte Carlo (MC) algorithms to draw samples from the posterior $\{\vec{\theta}_{i}\sim p(\cdot|\vec{D}):i=1,\ldots, n\}$: Markov chain Monte Carlo (see e.g.,~\cite{Sharma:2017wfu,Hogg:2017akh} for introductions in the context of physics) and nested sampling (NS;~\cite{Skilling04,skilling2006,Ashton:2022grj,Buchner_2023}). The samples facilitate parameter inference through estimates of the integrals in \cref{eq:marginal,eq:expectation}. Marginals may be estimated with histograms and expectations may be estimated through averages,
\begin{equation}
    \bar g = \frac1n \sum_{i=1}^n g(\vec{\theta}_{i}),
\end{equation}
however, in practice most algorithms (including NS) generate weighted samples, for which expectations are estimated through weighted averages,
\begin{equation}
\bar g = \frac{\sum_{i=1}^n  g(\vec{\theta}_{i}) w_i}{\sum_{i=1}^n w_i},
\end{equation}
where $w_i$ are the sampling weights.

Whilst  both MCMC and NS allow for parameter inference, their goals and thus their performance characteristics are somewhat different. NS was designed to allow one to make statistical estimates of compression, which appears in Bayesian model selection through the Bayesian evidence and in frequentist computation~\cite{Fowlie:2021gmr}. MCMC, on the other hand, allows for parameter inference without ever estimating compression. Computing compression ratios is fundamentally more computationally challenging, so it is generally accepted that NS algorithms command a higher computational cost.

MCMC runs are typically divided into two parts: a warm-up phase, that reduces bias, and a sampling phase, that eliminates noise~\cite{2023arXiv231102726M}. In the sampling phase, new posterior samples can be continually generated at a fixed computational cost. NS runs, on the other hand, compress from the prior to the bulk of the posterior mass and beyond. Consequently, new posterior samples cannot be generated at a fixed computational cost by running NS for longer and incomplete runs are of limited value, suffering from both bias (a truncation error) and noise. New NS runs, on the other hand, would waste computational effort compressing to the bulk of the posterior mass. Dynamic NS~\cite{2019S&C....29..891H} solves this problem by rewinding a completed NS run back to the bulk of the posterior mass and starting an independent run from that point. This amortizes the cost of compressing to the bulk of the posterior.

It should be noted therefore that NS algorithms are generally understood to be considerably more expensive than MCMC algorithms as a method for generating posterior samples, and that very little will beat a well-tuned MCMC. This is because NS algorithms are designed to compute the evidence, which requires compressing from prior through posterior and onto the peak of the likelihood. MCMC algorithms on the other hand cannot compute the evidence alone~(as they do not estimate compression), although other methods exist to augment these such as \code{MCEvidence}~\cite{Heavens:2017afc}, \code{FloZ}~\cite{Srinivasan:2024uax} or \code{harmonic}~\cite{Polanska:2024arc}. In practice, NS is used as a posterior sampler when the problem is multimodal, which can cause challenges for MCMC algorithms. In this paper, we demonstrate that modern MCMC algorithms can now be effective on multimodal problems, bringing the efficiency of MCMC to bear on problems previously only NS could tackle.

Among our total list of algorithms are two significant variants to the MCMC and NS paradigms. First, we consider an adaptive MCMC algorithm~(see, e.g.,~\cite{andrieu2008tutorial}).
In this algorithm, a diffusion model proposal~\cite{Hunt-Smith:2023ccp} is continually tuned in both the warm-up and sampling phases, and thus the efficiency continually increases towards one over the course of a run. Consequently, new posterior samples can be generated at an ever-decreasing computational cost. Tuning during the sampling phase could, however, introduce bias. Second, we consider an importance sampling method, based on an NS sampling scheme.

\subsection{Markov chain Monte Carlo}

Markov chain Monte Carlo (MCMC) algorithms have been used for decades within cosmology (see e.g.,~\cite{Christensen:2001gj,Lewis:2002ah}) and particle physics (see e.g.,~\cite{Baltz:2004aw,RuizdeAustri:2006iwb}) to explore complex posteriors, and can still be considered the state of the art in some cases. 
%(e.g.~inferring the parameters of the $\Lambda$CDM cosmological model~\cite{EXAMPLE_MCMC_COSMO}{\bf TODO: missing reference?}). 
In MCMC, we sequentially draw a set of samples $\{\vec{X_i}\}$ from the unnormalised posterior distribution. One of the benefits of MCMC is that the normalisation constant, or Bayesian evidence, need not be known. While this is helpful for parameter inference, it hinders model comparisons, which are typically performed using the Bayesian evidence.

Each $\vec{X_i}$ is a $D$-dimensional vector of numbers that represent each of the parameters in our $D$-dimensional parameter space, and the drawing of samples is a \emph{stochastic process}. Such a process is said to have the \emph{Markov property} if the conditional probability distribution of future states of the process (conditional on both past and present states) is only dependent on the present state. Our sampling process will have the Markov property if the conditional probability of the next sample taking a certain value depends only on the present sample, and not on all the samples that preceded it. A Markov chain Monte Carlo algorithm, therefore, is an algorithm that builds this feature into the selection of the samples. Performant MCMC algorithms generally have an adaptive element to tackle \textit{a priori} unseen problems, which technically breaks this Markov property, although providing the adaptivity is slow compared to the sampling timescale (or if this tuning period is discarded) this is rarely a problem in practice.

One of the simplest MCMC algorithms is the \emph{Metropolis-Hastings Algorithm}, which we'll use as an illustrative example. First, assume that we are at a point in the parameter space where the parameter values lie within the range of the prior. We will label this as the $t$-th point in the chain, with a parameter vector $\vec{X^{(t)}}$. We then define and use a \emph{proposal density} $Q(\vec{X'};\vec{X^{(t)}})$ to obtain a new sample $\vec{X'}$. As expected, this depends only on the current state of the system. It can be any fixed density, with support encompassing the posterior distribution support, from which it is possible to draw samples; in particular, it is not necessary for $Q(\vec{X'};\vec{X^{(t)}})$ to resemble the posterior that we are sampling in order to be useful. It is common to choose a simple, easily-sampled distribution, such as a multidimensional Gaussian.

We then choose whether to accept the new sample $\vec{X'}$ based on its acceptance ratio:
\begin{equation}
  \label{eq:metropolis}
a_{\vec{X'}}=\frac{p^*(\vec{X'}\given\vec{D})Q(\vec{X^{(t)}}\given\vec{X'})}{p^*(\vec{X^{(t)}}\given\vec{D})Q(\vec{X'};\vec{X^{(t)}})},
\end{equation}
where $p^*(\vec{X_i}\given\vec{D})$ is the unnormalised posterior for the parameter choice $\vec{X_i}$. If $a_{\vec{X'}} \ge 1$ the new state is accepted, otherwise we draw an i.i.d. random number $\mathcal{U}[0,1]$, and accept the proposed state if it is less than $a_{\vec{X'}}$. If we accept the new sample, it becomes the next point in our Markov chain. Otherwise, we stay at the current point and draw another proposal. Note, if $Q$ is a symmetric function of $\vec{X^{(t)}}$ and $\vec{X'}$, then the ratio of $Q$ factors disappears from $a_{\vec{X'}}$, and the Metropolis-Hastings method reduces to the Metropolis method, which involves a simple comparison of the posterior at the two candidate points in the Markov chain.

It can be shown that the probability distribution of the collection of accepted proposals $\{\vec{X^{(n)}}:n=1..t\}$ tends to the posterior distribution as $t\to \infty$, provided that $Q$ is chosen such that $Q(\vec{X'};\vec{X})>0$ for all $\vec{X},\vec{X'}$. Thus, by choosing points via the Metropolis-Hastings algorithm, we obtain samples from the unnormalised posterior.

Note that the presence of the caveat $t \to \infty$ implies that there is an issue of convergence in the application of the Metropolis algorithm, and this is to be expected from the Markov Chain nature of the method. Each element in the sequence $\vec{X^{(t)}}$ has a probability distribution that is dependent on the previous value $\vec{X^{(t-1)}}$ and hence, since successive samples are correlated with each other, the Markov Chain must be run for a certain length of time in order to generate samples that are effectively independent. The exact details of convergence depend on the particular posterior being sampled, and on the details of $Q$, and hence there is some element of tuning involved in getting the algorithm to run efficiently. 

The purpose of the definition in \cref{eq:metropolis} is to ensure that the Markov Chain used in the Metropolis method is reversible. By this it is meant that the chain satisfies the principle of detailed balance, i.e.~that the probability $T(\vec{X^a} ; \vec{X^b})$ of the chain making a transition from a state $\vec{X^a}$ to a state $\vec{X^b}$ is related to the reverse transition via $T(\vec{X^a}; \vec{X^b})P(\vec{X^a}) = T(\vec{X^b};\vec{X^a})P(\vec{X^b})$. This property is fundamental to how Metropolis sampling ensures convergence to the posterior.

For all MCMC algorithms, we must choose a point at which to start. Ideally, we would start from a draw from the target, though if we could draw from the target, we wouldn't be running MCMC in the first place. A common technique for choosing a starting point is burn-in or warmup. Here, one makes a preliminary choice of starting point and runs the chain for a number of `burn-in' iterations. The states of the chain in this period are discarded, and the position of the chain at the end of burn-in is used as the starting point. Other possibilities include draws from the prior and the mode of the target.

For a fair comparison of the MCMC techniques used in this paper, we use a common set of starting points for all MCMC chains and the same length of burn-in. 

\subsubsection{List of MCMC algorithms}

The efficiency of an MCMC sampler depends on the auto-correlation time of its chain, which can be reduced by a well-chosen proposal distribution. Thus, the key to an efficient MCMC implementation lies in constructing proposals judiciously. We consider the following MCMC codes:
 \begin{itemize}
     \item \verb|emcee|~\cite{2013PASP..125..306F,2019JOSS....4.1864F}: affine-invariant proposals are constructed from an ensemble of chains~\cite{2010CAMCS...5...65G}. The affine-invariance makes the algorithm self-tuning and amenable to problems with degeneracies and anisotropies.
     
     \item \verb|ptemcee|~\cite{2016MNRAS.455.1919V}: builds upon \verb|emcee| by parallel tempering among several ensembles of chains~\cite{PhysRevLett.57.2607,geyer1994estimating,2005PCCP....7.3910E}. Tempering involves raising the likelihood $\mathcal{L}^\beta$ to the power of a ``temperature'' $\beta$, in analogy with the Boltzmann factor where $\log\mathcal{L}$ plays the part of the energy, with ``hot'' being $\beta\to0$ and ``cold'' where $\beta\to 1$. An ensemble of cold chains targets the original posterior and ensembles of warm chains target tempered versions of the posterior. The tempered posteriors are more amenable to MCMC because modes are somewhat flattened. A special move swaps states between cold and warm chains. Consequently, the cold chains mix more quickly between modes. Whilst this can help in multimodal problems, abrupt changes in the tempered posterior as we change temperature (phase transitions) could cause problems. 
     
     \item \verb|zeus|~\cite{Karamanis:2021tsx}: an ensemble slice sampler~\cite{Neal2003,Karamanis:2020zss}. Proposals are based on slices of the posterior, potentially leading to big steps and thus small auto-correlation. Anisotropic and multimodal problems are tackled using the ensemble paradigm: affine-invariant slices are constructed from an ensemble of walkers and moves between modes are enabled by slices from one mode to another.
     
     \item \verb|MCMC-diffusion|~\cite{Hunt-Smith:2023ccp}: a mixture of local MH moves and global moves from a diffusion model trained on previous states of the chain. 
 \end{itemize}
See Table~\ref{tab:walkers} for a list of the number of walkers used for each test problem. We checked that the number of walkers chosen for each algorithm corresponded to the best available results given the computational constraints for each problem.

%%%%%%%%%%%%%%%%%%%%%%%%%%%%%
\subsection{Nested Sampling}
%%%%%%%%%%%%%%%%%%%%%%%%%%%%%%

Nested sampling (NS) was introduced by John Skilling~\cite{skilling2006} as a general method of Lebesgue integration, with the main aim of efficiently evaluating Bayesian evidence,
\begin{equation}
Z = \int_\Sigma \mathcal{L}(\vec\theta) \, p(\vec\theta) \,\mathrm{d} \theta.  \label{eq:Z}
\end{equation}
A naive approach to evaluating \cref{eq:Z} using standard quadrature quickly fails as the dimension of $\vec\theta$ grows, as well as when much of the likelihood is contained in small regions of the prior volume.
Both properties are typical of most physical modelling problems. Simulated-annealing MC methods offer an alternative approach to evaluation of the Bayesian evidence~\cite{albert2015simulated}; however, as was shown in~\cite{skilling2006}, simulated annealing is fundamentally unable to handle likelihoods with phase transitions, which are also commonplace in physical models.

We will now briefly describe the basic NS algorithm, which is, at its core, a compression algorithm. 
Given any particular likelihood value, $\lambda$, it is possible to define a corresponding prior volume, $X(\lambda)$, such that $X(\lambda) = \int_{\mathcal{L}(\theta) > \lambda} p(\theta) d\theta$. The essence of NS is to transform the multidimensional integral of \cref{eq:Z} into a one-dimensional integral over $X$,
\begin{align}
    Z = \int \mathcal{L}(X) \,\mathrm{d} X.  \label{eq:Z_X}
\end{align}
This transformation simplifies the problem significantly, as it reduces the dependence on the dimensionality of $\theta$.

This transformation changes the problem into estimating how much prior volume, $X(\lambda)$ to place at each iso-likelihood contour, $\lambda$, resulting in a so-called shrinkage process of estimation.
The algorithm starts at $X(\lambda_0 \triangleq 0) \triangleq 1$, where we have precise knowledge of the amount of enclosed prior volume, and marches inwards to higher likelihood values.
A set of $\nlive$ samples are uniformly drawn from the likelihood constrained prior distribution, $\{\theta_n : n=1\ldots\nlive, \mathcal{L}(\theta_n) > \lambda_0\}$. This set is called the set of `live points'.
The sample among the live points with the lowest likelihood value, $\lambda_i$, is selected and replaced with a new sample uniformly drawn within the new likelihood $\lambda_i$. This procedure generates an ordered set of likelihood values, samples, and associated prior volumes, $\mathcal{S} = \{(\lambda_i, \theta_i, X_i): i=1..S\}$, which provide a means to estimate $Z$ via numerical integration,
\begin{align}
    Z \approx \sum_i \lambda_i \Delta X_i. \label{eq:Z_approx}
\end{align}
The key is that when selecting the lowest likelihood point from the live set, there is a known distribution for $X_i$.
By definition, it is given by the order statistic of the distribution of all $X$ in the live set.
Since we carefully ensure that each live point is \textit{uniformly} drawn from within the given likelihood contour, the order statistic is simply given by the Beta distribution.
The algorithm proceeds until an implementation-specific stopping criterion is met, usually based on an estimation of the remaining evidence in the live points.
Since its introduction there have been numerous explorations and improvements to the basic NS algorithm, but the core concept of compression remains in all variations~\cite{Ashton:2022grj,Buchner_2023}.

One of the key features of NS is its ability to explore complex, multi-modal distributions effectively. This is achieved through the use of the set of live points which allow the algorithm to sample from different modes of the distribution concurrently. This feature, combined with the reduction in dimensionality of the problem, makes NS particularly well-suited for Bayesian inference in high-dimensional spaces, where traditional methods like MCMC may struggle.
Furthermore, NS provides an elegant solution to the issue of phase transitions in likelihood surfaces. By athermally contracting the prior volume and sampling from the constrained space, NS adaptively focuses on regions of interest, seamlessly navigating across different phases of the likelihood landscape.

A by-product of the numerical integration performed by \cref{eq:Z_approx} is a means to compute the posterior distribution, or rather, samples from it.
By comparing \cref{eq:Z} with \cref{eq:Z_approx} we can see that the transformation to a one-dimensional enclosed prior volume implies that the set of samples $\mathcal{S}$ \textit{is} the posterior with a reweighting of the individual samples.
The relative weight of sample $i$ is given by,
\begin{align}
    w_i \triangleq \lambda_i \Delta X_i \label{eq:Z_weight}.
\end{align}
In both \cref{eq:Z_approx,eq:Z_weight} the $\Delta X_i$ are random variables due to the fact that the shrinkage process is a sampling procedure. Therefore, expectations are taken using the known properties of the order statistic distribution.
Using the above weights, $w_i$, it is then simple to resample the posterior to uniform weights and use the samples just as you would use samples from MCMC, which also produces uniformly weighted samples.

%%%%%%%%%%%%%%%%%%%%%%%%%%%%%%%%%%%%%%%%%
\subsubsection{List of NS algorithms}
%%%%%%%%%%%%%%%%%%%%%%%%%%%%%%%%%%%%%%%%%%

We compared several Nested Sampling (NS) algorithms implemented in \code{Bilby}~\cite{Ashton:2018jfp}, including \code{MultiNest}~\cite{0809.3437,1306.2144}, \code{PolyChord}~\cite{Handley:2015}, \code{Dynesty}~\cite{Speagle:2019ivv},  \code{UltraNest/MLFriends}~\cite{2016S&C....26..383B,2019PASP..131j8005B}, and \code{Nessai}~\cite{Williams:2023ppp}. We also integrated \code{JaxNS}~\cite{2020arXiv201215286A} and \code{Nautilus}~\cite{lange2023nautilus} into \code{Bilby}. Here is a summary of their methodologies and characteristics:

\begin{itemize}
    \item \textbf{\code{MultiNest}}: The original NS algorithm, using ellipsoidal rejection sampling to bound the likelihood contour. We used the Python interface provided by \code{PyMultiNest}.
    \item \textbf{\code{PolyChord}}: Utilizes 1D slice sampling and clustering analysis to handle separable posterior modes. We explored the default settings, focusing on its ability to step out in anisotropic problems and navigate multimodal distributions. We used the Python interface provided by \code{PyPolychord}.
    
    \item \textbf{\code{Dynesty}}: Offers a variety of rejection and step sampling methods, including dynamic NS. We used the default sampling method, which in our case was random walk sampling away from a current live point since our test problems were all between 10D and 20D. The only exception was the 5D $\Lambda$CDM physics example (see~\cref{subsec:CMB}), for which uniform sampling is the default method since this problem is less than 10D.
    
    \item \textbf{\code{UltraNest/MLFriends}}: Implements various region and step sampling methods, along with reactive NS, a generalization of dynamic NS. Both reactive and static nested samplers were explored. The MLFriends version of UltraNest was used in this work.
    
    \item \textbf{\code{Nessai}}: Combines NS with artificial intelligence (AI) by using normalizing flows for sampling from the constrained prior.
    
    \item \textbf{\code{JaxNS}}: A probabilistic programming framework that implements NS in JAX for high performance through compilation. It employs 1D slice sampling without a step out procedure, using exponential contraction from the bounds of the prior domain.
    
    \item \textbf{\code{Nautilus}}: A pure-Python package for posterior and evidence estimation, utilizing importance sampling based on NS and neural networks.
\end{itemize}

Each algorithm's unique approach and specific features are designed to optimize performance and accuracy in NS, catering to the diverse needs of astrophysical research and beyond. 
Broadly they divide into two categories~\cite{Ashton:2022grj}: region(/rejection) samplers and path(/step) samplers. Region samplers aim to construct an explicit proxy for the likelihood-constrained prior distribution, and sample from that. They will be highly performant in low dimensions, but suffer from a curse of dimensionality in high dimensions, becoming exponentially inefficient above some problem- and sampler-dependent dimensionality $d_0\sim\mathcal{O}(10)$. \code{MultiNest}, \code{Nessai}, \code{Nautilus} and \code{MLFriends} fall into this category. Path samplers run a Markov chain starting from a random live point until it has ``forgotten'' where it started. These are less efficient in low dimensions, but their inefficiency scales linearly with dimensions, allowing them to operate in principle in hundreds of dimensions. The state-of-the-art is achieved with slice sampling~\cite{2022PSFor...5...46B}, as implemented in \code{PolyChord} and \code{JaxNS} (although \code{JaxNS} simultaneously uses an ellipsoidal element), with \code{Dynesty} and \code{UltraNest} switching to slice-sampling modes in high dimensions.

For each NS implementation we explored over a small grid of hyperparameters.
In selecting the explored hyperparameters for each implementation we tried to choose similarly to how a typical end-user would in their analysis.
This `end-user mindset' reflects the overall intent of the paper to help the community chose most wisely for their analysis. 
To this end, we chose to explore the most well-documented hyperparameters specific to each implementation, and used each implementation's default stopping conditions. 

Across most implementations there was a concept similar to the number of live points, in which case we explored settings of $\nlive=\{25D, 50D, 100D\}$.
We note that in the modern era most NS implementations do not have the concept of a static number of live points, but rather treat it as a dynamic quantity (particularly for the purposes of parallelisation). 
In these cases, it was often possible to set the notion of a minimal number of live points.
By far the most impactful hyperparameter for the quality and computation expense of each run is the number of live points.

\section{Methodology} \label{sec:methodology}

In this section, we provide a brief outline of how we chose to compare samplers. 

\subsection{Evaluating Nested Sampling implementations}

In this study, we aimed to evaluate different NS implementations to find the most computationally efficient solution for various problems, adopting the mindset of the `typical experimenter'. 
The goal was to identify an NS implementation that achieves convergence with minimal computational effort, facilitating exploration within limited time frames. 
Therefore, our approach was to explore each NS implementation within consistent hyperparameter spaces constrained to a fixed computational limit per run, and to identify the most economical converging run. 

Since each NS implementation has differing algorithmic structure, ensuring consistent hyperparameter spaces was reduced to a simple scheme of controlling the most well-documented hyperparameters. 
Specifically, for implementations with the concept of live points, the exploration space was $\{25D, 50D, 100D\}$, except for the Rastrigin problem which also included $\{200D, 500D, 1000D\}$. An exception to this hyperparameter search was the \texttt{nautilus} algorithm, for which the number of live points was consistently set to the default value of 2000 as per the algorithm author's advice.
For implementations using Markov Chains, $5D$ accepted proposals per sample were selected.
Where multiple modes of operation existed, e.g. the Static or Reactive mode of \texttt{ultranest}, the top recommended mode(s) were used. 
Static nested sampling was chosen over dynamic to maintain consistency in the number of live points. 
All other hyperparameters were left at their default, including stopping conditions.
Each run was limited to one hour, except for the Rastrigin problem, which had a two-hour limit, reflecting practical constraints faced by non-expert experimenters.

Convergence was assessed via the Wasserstein distance measure, also known as the Earth-Mover's distance from a reference posterior. 
First, a reference posterior we generated for each problem using a high-resolution run of NS, making sure that this reference posterior closely matched the true posterior as closely as possible for each problem. Specifically, we used \texttt{JaxNS} with $2000 D$ live points, $10 D$ slices per acceptance.
Then, an upper bound on the multi-variate Wasserstein-$1$ distance was used as the distance measure, namely $W_1(P, Q) \leq \sum_i W_1(P_i, Q_i)$ where $W_1$ is the 1-norm Wasserstein distance, and $P_i, Q_i$ are the respective one-dimensional marginals of the distributions.
We then used per-problem convergence thresholds to identify convergence. 
We opted to compute the Wasserstein distance in the unit-cube scaled parameter spaces, hence the commonly used fiducial threshold of 0.1 was used for all problems except Eggbox and Rastrigin, which assumed 0.3 and 0.2 respectively, due to the more challenging nature of those problems.
Plots of the Wasserstein upper bound for the cheapest converged NS runs are shown in Figure~\ref{fig:wasserstein_upper}.
Finally, for each problem and NS implementation, the cheapest run within the convergence threshold was identified as the cheapest converging run.
During the course of our experimentation, we extended the hyperparameter space and computational limits for the Rastrigin problem, so that at least one NS implementation converged for each case.

\subsection{Diagnostics \& Metrics}\label{sec:metrics}

We are using two categories of algorithm to sample from a posterior distribution: MCMC and NS. In each case, we want to assess convergence, diagnose any problems with runs and measure performance. Although under appropriate conditions MCMC algorithms may be proven to converge asymptotically, at some point we must stop our chain(s). We would like to know whether the samples obtained at that point are representative of the underlying stationary distribution. They will not be representative in a short chain if the chain explores the posterior distribution slowly. This `slow mixing' may occur, for example, in multi-modal problems.

Unfortunately, assessing the convergence of an MCMC chain is extremely challenging~\cite{doi:10.1080/01621459.1996.10476956,2023arXiv231102726M} and arguably we at best make checks of pseudo-convergence, since no checks can tell us what would happen if we were to run the chain for longer~\cite{geyer2011introduction}. We use \code{arviz}~\cite{arviz_2019} to estimate the Gelman-Rubin $\rhat$ parameter~\cite{10.2307/2246093,doi:10.1080/10618600.1998.10474787} for each model parameter. This requires more than one MCMC chain, ideally initialised at different starting points, as it compares the within-chain and between-chain variances of a parameter. See~\cite{2019arXiv190308008V,2021arXiv211013017M} for up-to-date discussions about the limitations, interpretation and developments of the $\rhat$ parameter. Recently, $\rhat \le 1.01$ was recommended as a convergence check~\cite{2019arXiv190308008V}.

Ensemble MCMC samplers are somewhat problematic with regard to diagnostics and metrics. In ensemble samplers, efficient proposals are constructed using an ensemble of walkers. As the proposals are constructed from the ensemble, the walkers are not independent chains. Consequently, the conventional interpretations of and rules of thumb about the $\rhat$ parameter (and to some degree the ESS estimates discussed later) are open to question.

NS does not have the same question of convergence to a stationary distribution, since it is not building a Markov chain. This may, however, be seen as a drawback as we cannot efficiently generate more samples by running NS for longer (though see the discussion on dynamic NS). In NS, rather, the key questions are whether we prematurely terminated NS and whether we genuinely drew independent samples from the constrained prior because if we did not, the NS results are faulty. In~\cite{Fowlie:2020mzs}, a new test was introduced to diagnose problems in this aspect of NS runs. The diagnostic, $p_{\rm insert}$, makes use of the fact that the rank indices of new live points should be uniformly distributed. The result of this diagnostic is a $p_{\rm insert}$ from a one-sample test between the rank indices from an NS run and a uniform distribution. This test has not yet been implemented across all the algorithms used in this paper, so there are some missing $p_{\rm insert}$ in certain test problems. As always, care should be taken in interpreting $p_{\rm insert}$, though small $p_{\rm insert}$ could indicate a failure to sample from the constrained prior and thus a faulty NS run.\footnote{If we fail to faithfully sample from the constrained prior whilst compressing to the posterior bulk, it might impact the evidence but not the posterior. If, on the other hand, it happens whilst compressing through the bulk of the posterior, posterior inferences may be faulty as well.}

To quantify the performance of an algorithm, we consider what we obtained and for what cost. As we are concerned with sampling from the posterior, we measure what we obtained by the effective number of independent samples drawn from the posterior, $\neff$. Unfortunately, the samples we obtain are not independent draws from the posterior. In Markov chains the samples are correlated as the state of the chain depends on the previous state. In NS, the samples obtained are weighted, as they are weighted draws from the NS procedure rather than draws directly from the posterior. The cost that we are concerned with is the number of likelihood evaluations, $\nlike$, such that as a metric we consider
\begin{equation}
\text{Efficiency} = \frac{\neff}{\nlike}.
\end{equation}
This doesn't take into account the fact that different algorithms may be parallelized more efficiently than others. The effective sample size may depend on the parameter or quantity of interest; in such cases we take the minimum among model parameters. See \cref{sec:ess} for a discussion of effective sample size estimates in MCMC and NS. Broadly speaking, $\neff$ estimates may not reliable when $\neff \lesssim 100$.

\section{Results}\label{sec:results}

We investigated the algorithms in \cref{sec:algorithms} by running them on several test functions and on examples inspired by global fits in particle physics and cosmology. The test functions were chosen to exhibit pathological features, including high-dimensionality, multi-modality, degeneracy, and phase transitions.

\subsection{Eggbox}

\begin{figure} [h]
\centering
\includegraphics[scale = 0.7]{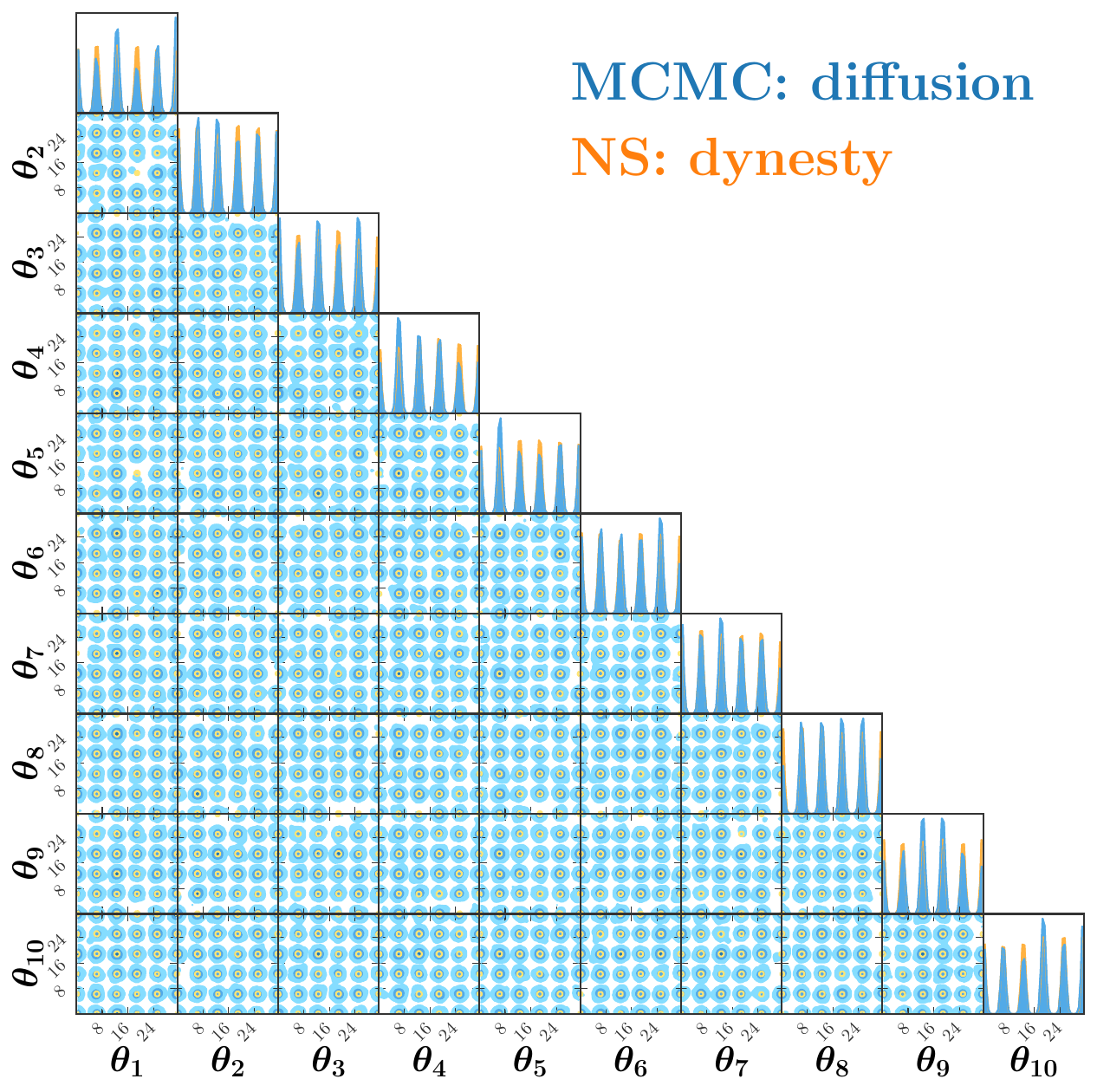}
\caption{Corner plot of the $10D$ Eggbox function in \cref{eq:eggbox}. The plot shows the posteriors from the best performing MCMC and NS implementations.}
\label{fig:eggbox}
\end{figure}

The Eggbox function, given by
\begin{equation}\label{eq:eggbox}
    f(\bm{\theta}) = \left(2 + \prod_{i = 1}^D \left[\cos\left(\frac{\theta_i}{2}\right)\right]\right)^{5},
\end{equation}

leads to a complicated multimodal posterior distribution, as illustrated in a 10-dimensional corner plot shown in \cref{fig:eggbox}. The posterior mean obtained through different algorithms is summarised in \cref{tab:Eggbox_means}, while \cref{tab:Eggbox_metrics} presents their corresponding performance metrics and diagnostics.\footnote{We note that for a highly multi-modal problem, the posterior mean proves to be a statistic which is sensitive to misfitting, but otherwise is not particularly useful in comparison with for example the individual mode means.}  The term ``Modes Found'' refers to the percentage of modes for which an algorithm generates sufficient samples to outline a $1\sigma$ contour within the 2D marginal distributions in \cref{fig:eggbox}. We note that this is a considerably easier problem then enumerating the $5^{10}\approx10^7$ modes present, and is only marginally interesting for a distribution with a high degree of axis-aligned symmetry such as the Eggbox.

Scrutinising these results shows that \verb|MCMC-diffusion| and \verb|ptemcee| demonstrate superior performance in locating most of the modes efficiently compared to NS methods.  Specifically, \verb|dynesty| excels in identifying the highest number of modes among the algorithms, although it requires significantly more likelihood evaluations than MCMC approaches to achieve this outcome.  \verb|MCMC-diffusion|'s efficacy in navigating multimodal distributions, characterised by its ability to transition between modes efficiently, is evident in this scenario.  While \verb|ptemcee| performs well, it struggles with modes situated at the parameter space's edges, leading to narrower posterior standard deviations.

Notably, other traditional MCMC algorithms fail to converge within the specified computational constraints. Regarding \verb|MCMC-diffusion|, multiple chains are executed to identify various modes efficiently, as outlined in the original paper~\cite{Hunt-Smith:2023ccp}. This allows the algorithm to overcome the traditional limitations of MCMC (e.g., fixed step size, local optima trapping). It should be noted that in a system of high symmetry (namely equal mode weighting and separation) multi-start MCMC produces the correct result. However, if modes are asymmetric, multi-start MCMC will initially populate the modes in proportion to the size of their burn-in basin of attraction, which may not be the same as their posterior weight. Only after running for long enough to transition between modes will the correct posterior be sampled, which may be after the computational budget has been exhausted, and often long after a convergence criterion has been triggered. We shall see this for the spike-slab function discussed in \cref{sec:spike_and_slab}.

\begin{table}[]
    \centering
    \begin{tabular}{cccccc}
    \toprule
    Algorithm & Diagnostic & \neff & \nlike & Efficiency & Modes Found\\
    \midrule
    MCMC: \verb|emcee| & \rhat = 1.07 & 31 & 1.50M & 2.06 $\times 10^{-5}$ & 69\%\\
    \midrule
    MCMC: \verb|ptemcee| & \rhat = 1.00 & 110k & 1.50M & 0.0733 & 91\%\\
    \midrule
    MCMC: \verb|zeus| & \rhat = 1.10 & 25 & 1.50M & 8.80 $\times 10^{-6}$ & 53\%\\
    \midrule
    \rowcolor{row-highlight}
    MCMC: \verb|MCMC-diffusion| & \rhat = 1.00 & 120k & 1.08M & 0.111 & 93\%\\
    \midrule
    \rowcolor{row-highlight}
    NS: \verb|dynesty| & $p_{\rm insert}$ = 0.23 & 2.04k & 26.8M & $7.59\times 10^{-5}$ & 96\%\\
    \bottomrule
    \end{tabular}
    \caption{Algorithm performance metrics for the $10D$ Eggbox test function.}
    \label{tab:Eggbox_metrics}
\end{table}

%%%%%%%%%%%%%%%%%%%%%%%%%%%
\subsubsection{Rosenbrock}
%%%%%%%%%%%%%%%%%%%%%%%%%%%

\begin{figure} [h]
\centering
\includegraphics[scale = 0.7]{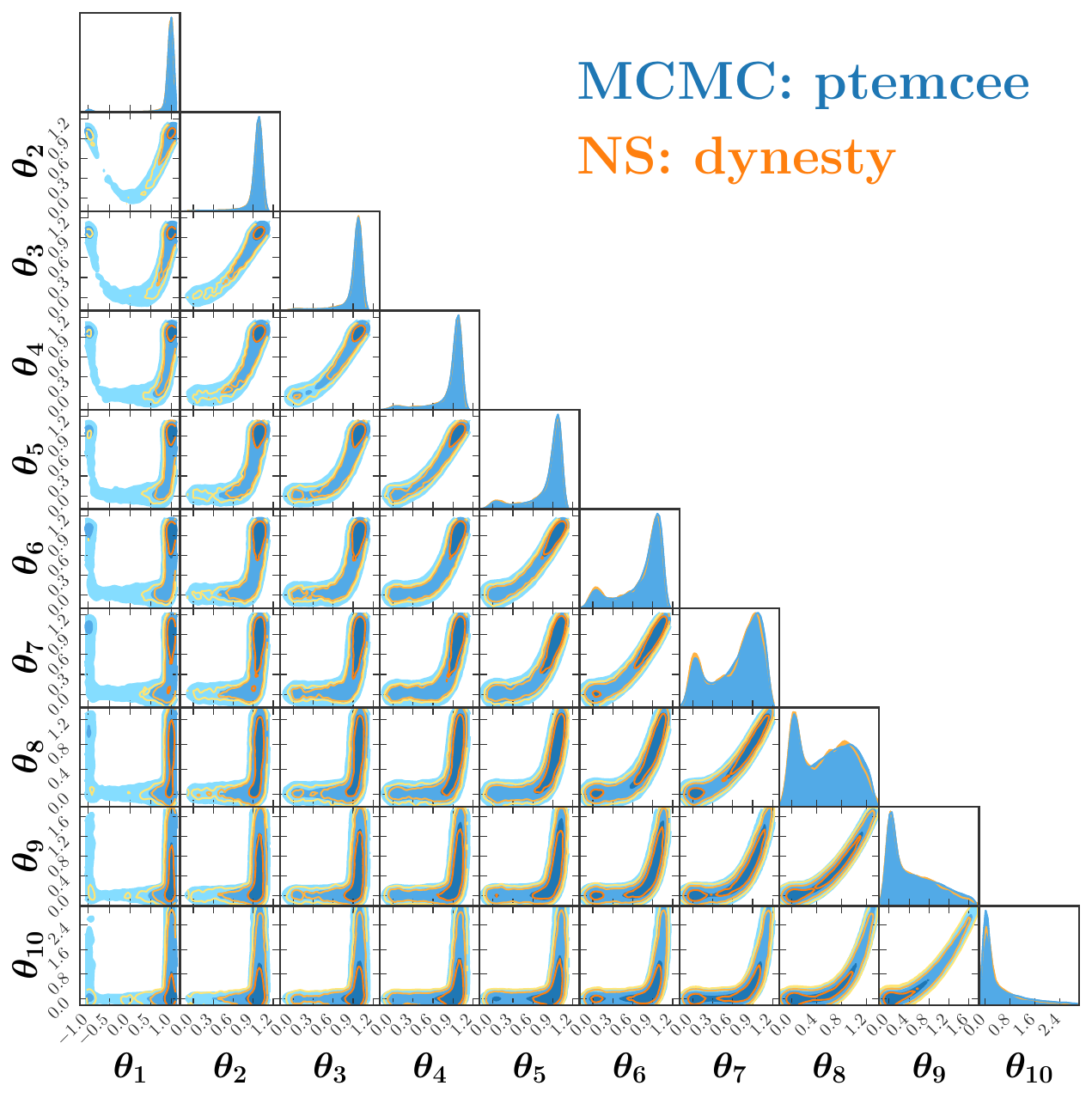}
\caption{Corner plot of the $10D$ Rosenbrock function.}
\label{fig:Rosenbrock}
\end{figure}

The Rosenbrock function~\cite{Rosenbrock1960}, defined as
\begin{equation}
    f(\bm{\theta}) = \sum_{i=1}^{D-1}[100 \times (\theta_{i + 1} - \theta_i)^2 + (1-\theta_i)^2],
\end{equation}
displays a banana-shaped degeneracy in two dimensions.  A 10-dimensional corner plot illustrating this function is depicted in \cref{fig:Rosenbrock}, accompanied by a posterior mean detailed in \cref{tab:Rosenbrock_means} and summary metrics provided in \cref{tab:Rosenbrock_metrics}.

Analyzing the outcomes, the following key observations arise.  Among the NS algorithms, \verb|dynesty| appears to perform more accurately compared to others, as indicated by the posterior mean.  The other NS algorithms exhibit a slight underestimation in parameters $\theta_4-\theta_{10}$, with narrower standard deviations.  Further examination of the posterior plots reveals that the remaining NS algorithms fail to explore the entire distribution width effectively.  Notably, the second peak in parameters $\theta_6-\theta_8$ is notably shifted to the left, suggesting that these algorithms are unable to locate the broader peak positioned further to the right.

Algorithms such as \verb|emcee| and \verb|zeus| demonstrate insufficient convergence within the specified number of likelihood evaluations, as evidenced by the $\hat{R}$ metric for these methods.  With \verb|dynesty| requiring significantly more likelihood evaluations than \verb|ptemcee| or \verb|MCMC-diffusion|, the results imply that Markov chain Monte Carlo methods are once again demonstrating greater efficiency compared to NS in this context. 

\begin{table}[]
    \centering
    \begin{tabular}{ccccc}
    \toprule
    Algorithm & Diagnostic & \neff & \nlike & Efficiency\\
    \midrule
    MCMC: \verb|emcee| & \rhat = 1.35 & 7.10k & 4.00M & 0.00180\\
    \midrule
    \rowcolor{row-highlight}
    MCMC: \verb|ptemcee| & \rhat = 1.00 & 4.55k & 550k & 0.00828\\
    \midrule
    MCMC: \verb|zeus| & \rhat = 1.09 & 26.5k & 2.00M & 0.0131\\
    \midrule
    MCMC: \verb|MCMC-diffusion| & \rhat = 1.01 & 2.12k & 700k & 0.00303\\
    \midrule
    \midrule
    \rowcolor{row-highlight}
    NS: \verb|Dynesty| & $p_{\rm insert}$ = 0.38 & 2.6k & 5.1M & 5.1$\times 10^{-4}$ \\
    \midrule
    NS: \verb|Multinest| & $p_{\rm insert}$ = 0.1 & 2.6k & 167k & 0.02 \\
    \midrule
    NS: \verb|Polychord| & $p_{\rm insert}$ = 0.69 & 3k & 2.9M & 0.001 \\
    \midrule
    NS: \verb|Nessai| & $p_{\rm insert}$ = 5.13$\times 10^{-5}$ & 3.9k &  203k & 0.02 \\ 
    \midrule
    NS: \verb|JaxNS| &  & 1.6k & 3.9M & 4.07$\times 10^{-4}$ \\
    \midrule
    NS: \verb|Nautilus| &  & 10k & 179k & 0.056 \\
    \bottomrule
    \end{tabular}
    \caption{Summary of algorithm performance for $10D$ Rosenbrock test function. 
    }
    \label{tab:Rosenbrock_metrics}
\end{table}

%%%%%%%%%%%%%%%%%%%%%%%%%%
\subsubsection{Rastrigin}
%%%%%%%%%%%%%%%%%%%%%%%%%

The Rastrigin function~\cite{rastrigin1974systems} is given by
\begin{equation}
    f(\bm{\theta}) = 10 D + \sum_{i=1}^{D}\left[ \theta_i^2 - 10 \cos(2\pi\theta_i)\right],
\end{equation}
where $D$ is the number of dimensions of $\bf{\theta}$. 

\begin{figure} [h]
\centering
\includegraphics[scale = 0.7]{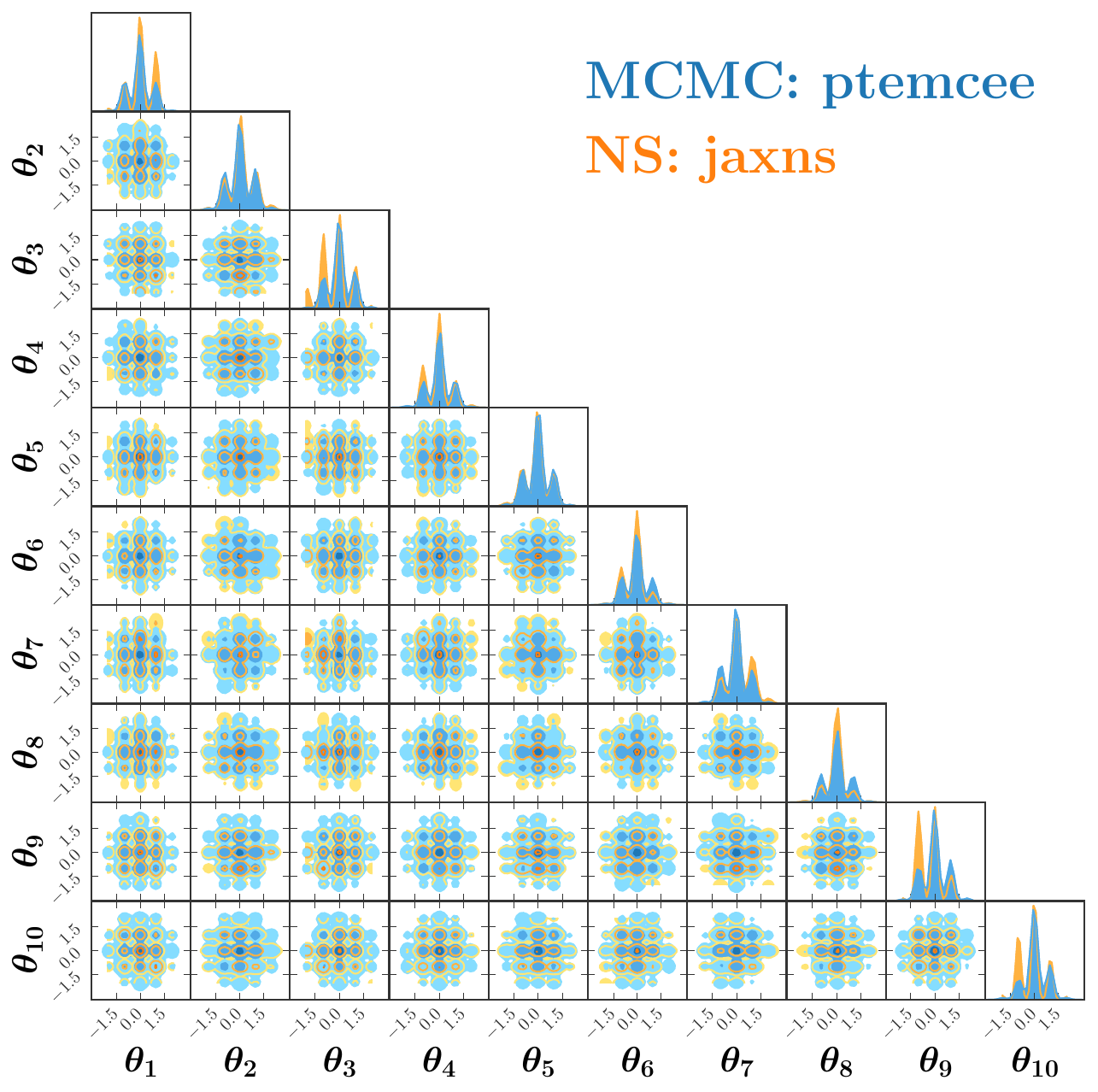}
\caption{Corner plot of the $10D$ Rastrigin function.}
\label{fig:Rastrigin}
\end{figure}

The Rastrigin function is also multimodal.  A 10-dimensional corner plot depicting this function is presented in \cref{fig:Rastrigin}, complemented by the computed posterior mean documented in \cref{tab:Rastrigin_means} and summarised metrics detailed in \cref{tab:Rastrigin_metrics}.

Key observations from the analysis of these results include the following.  The \verb|ptemcee| algorithm demonstrates a notable proximity to the true distribution compared to other methods, requiring a significantly lower number of likelihood evaluations (400,000) compared to \verb|Dynesty| (11M) to achieve this result.  This efficiency indicates a compelling performance advantage for \verb|ptemcee| in capturing the underlying distribution accurately within a more feasible computational budget.  The contrast in performance metrics highlights the varying computational demands and effectiveness of different algorithms when applied to this multimodal function.  

\begin{table}[]
    \centering
    \begin{tabular}{ccccc}
    \toprule
    Algorithm & Diagnostic & \neff & \nlike & Efficiency\\
    \midrule
        MCMC: \verb|emcee| & \rhat = 1.12 & 25 & 400k & 6.28 $\times 10^{-5}$\\
    \midrule
    \rowcolor{row-highlight}
    MCMC: \verb|ptemcee| & \rhat = 1.00 & 14.1k & 400k & 0.0353\\
    \midrule
    MCMC: \verb|zeus| & \rhat = 1.12 & 27 & 400k & 2.67 $\times 10^{-5}$\\
    \midrule
    MCMC: \verb|MCMC-diffusion| & \rhat = 1.01 & 11.2k & 400k & 0.0275\\
    \midrule
    \rowcolor{row-highlight}
    NS: \verb|JaxNS| &  & 54k & 104M & 5.2$\times 10^{-4}$ \\
    \bottomrule
    \end{tabular}
    \caption{Summary of algorithm performance for $10D$ Rastrigin test function.}
    \label{tab:Rastrigin_metrics}
\end{table}

%%%%%%%%%%%%%%%%%%%%%%%%%%%%%%%%%%%
\subsubsection{Spike-slab function}\label{sec:spike_and_slab}
%%%%%%%%%%%%%%%%%%%%%%%%%%%%%%%%%%%

The spike-slab function is commonly used in statistical modeling, particularly in the context of variable selection, as a Bayesian prior distribution.  It combines a point-like likelihood mass (spike) with a continuous distribution (slab).  This distribution is typically used to encourage sparsity in model parameters, allowing for automatic variable selection during model estimation.  

The spike-slab function may have a spike that is similar in volume to the slab which can pose challenges for Markov chain Monte Carlo algorithms.  These challenges include poor mixing between modes, leading to inefficient exploration, and convergence issues due to sharp transitions and narrow bridges.

\begin{figure} [h]
\centering
\includegraphics[scale = 0.7]{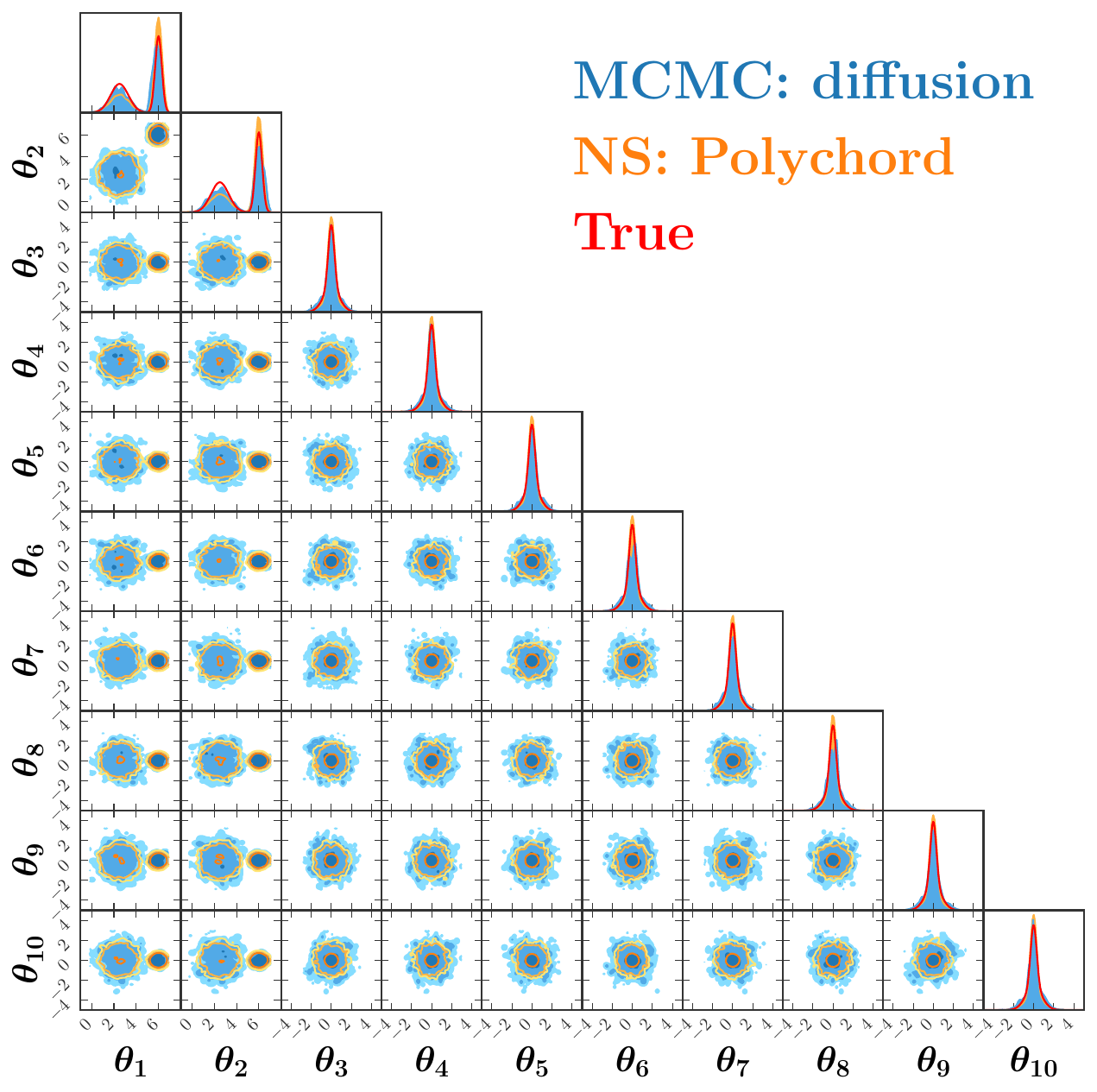}
\caption{Corner plot of the $10D$ spike-slab function.}
\label{fig:spikeslab}
\end{figure}

Our spike-slab function is the sum of a narrow $2D$ Gaussian and a broad multi-variate Gaussian. The narrow Gaussian has a mean of $6$ for each parameter and a diagonal covariance matrix with a standard deviation of $0.08$ for each parameter. Similarly, the broader Gaussian has a mean of $2.5$ and a standard deviation of $0.8$. A corner plot of this function in $10D$ is shown in \cref{fig:spikeslab}, along with the posterior mean and error in \cref{tab:spikeslab_means} and summary metrics in \cref{tab:spikeslab_metrics}.

Most MCMC algorithms, except for \verb|MCMC-diffusion|, struggle to correctly balance the weights assigned to the spike and slab components.  Despite indicating convergence based on the $\hat{R}$ metric, these algorithms converge to incorrect estimates, underscoring the complexity and potential confusion posed by the spike-slab function for MCMC methods. 
Even \verb|ptemcee| demonstrates a tendency to incorrectly estimate the relative weight of the spike and slab components, as can be seen by the underestimation of the posterior mean in parameters $\theta_1$ and $\theta_2$ (see table~\ref{tab:spikeslab_means}), indicating \verb|ptemcee| has focused too heavily on the slab. This further highlights the difficulty in achieving accurate balance when dealing with the spike-slab function within Bayesian inference settings. 

Nested sampling can also show similar difficulties. Whilst region samplers with a well-trained proxy will capture mode weighting up to some statistical error, chain based samplers on their own will mis-estimate modes that they cannot ergodically mix~\cite{Handley:2015}. Samplers with semi-independent mode tracking such as \code{PolyChord} and \code{GGNS}~\cite{2023arXiv231203911L} mitigate this (up to a statistical error which they track).
 This is backed up by both the posterior mean and Wasserstein distances, since \code{PolyChord} was the only NS algorithm that got close to the true distribution in this test problem.

 These observations emphasise the intricate nature of algorithmic performance, and are important to bear in mind when considering the challenges associated with complex model priors.

\begin{table}[]
    \centering
    \begin{tabular}{ccccc}
    \toprule
    Algorithm & Diagnostic & \neff & \nlike & Efficiency\\
    \midrule
    MCMC: \verb|emcee| & \rhat = 1.01 & 1.44k & 250k & 0.00600\\
    \midrule
    MCMC: \verb|ptemcee| & \rhat = 1.00 & 22k & 250k & 0.0880\\
    \midrule
    MCMC: \verb|zeus| & \rhat = 1.00 & 8.79k & 250k & 0.0293\\
    \midrule
    \rowcolor{row-highlight}
    MCMC: \verb|MCMC-diffusion| & \rhat = 1.00 & 9.66k & 90k & 0.107\\
    \midrule
    NS: \verb|Polychord| & $p_{\rm insert}$ = 0.32 & 15k & 7.6M & 0.002 \\
    \midrule
    \rowcolor{row-highlight}
    NS: \verb|JaxNS| &  & 5.6k  & 4M  & 0.001 \\
    \midrule
    \bottomrule
    \end{tabular}
    \caption{Summary of algorithm performance for $10D$ spike-slab test function}
    \label{tab:spikeslab_metrics}
\end{table}

%%%%%%%%%%%%%%%%%%%%%%%%%%%%%%%%%%
\subsection{Cosmology --- Planck}
\label{subsec:CMB}
%%%%%%%%%%%%%%%%%%%%%%%%%%%%%%%%

The Lambda Cold Dark Matter ($\Lambda$CDM) model stands as the prevailing cosmological paradigm, aiming to describe the universe's composition, structure, and evolution. At its core, the $\Lambda$CDM model is anchored in the Big Bang Theory and inflation, further augmented by the cosmological constant ($\Lambda$) and cold dark matter (CDM) components to address observations not fully explained by traditional cosmological frameworks.

The model's elegance lies in its foundation upon a mere six parameters, each playing a pivotal role in shaping our understanding of our Universe. Namely, the angular size of the sound horizon at the surface of last scattering ($\theta_s$), the baryon density ($\Omega_b h^2$), the dark matter density ($\Omega_c h^2$), the amplitude of primordial fluctuations ($A_s$), the scalar spectral index ($n_s$) and the optical depth to reionization ($\tau$).

Bayesian global fits of the $\Lambda$CDM model are the state of the art for extracting knowledge of these fundamental cosmological parameters. For any given parameter point, the
Boltzmann code CLASS~\cite{lesgourgues2011cosmic} can be used to compute various power spectra of the fluctuations in the cosmic microwave background, which can then be compared to current observations. The spectra are expressed in terms of the multipole moment $\ell$, related to angular scales on the sky, with the notation $C_\ell$ used for the angular power spectrum. This spectrum reflects the variance of the temperature fluctuations as a function of the angular scale. Depending on the nature of the fluctuations, we can categorize them into different types, primarily TT, TE, and EE, each offering unique insights into the early universe. The TT (Temperature-Temperature) spectrum represents the correlation of temperature fluctuations, providing a direct measurement of the cosmic microwave background's anisotropies. The TE (Temperature-E-mode Polarization) spectrum captures the cross-correlation between temperature fluctuations and the E-mode polarization pattern, offering a complementary view that helps break degeneracies in cosmological parameter estimation. Lastly, the EE (E-mode Polarization-E-mode Polarization) spectrum focuses on the correlation within the E-mode polarization itself, sensitive to the early universe's density fluctuations and the epoch of reionization. We denote them as $C_\ell^\text{TT}$, $C_\ell^\text{TE}$ and $C_\ell^\text{EE}$.  

With those we can compute the likelihood function (see Eq. 3.7 in~\cite{Cole_2022})
\begin{equation}
\ln p (\hat{C}(\theta)|C) = -\frac{1}{2} \sum_\ell (2\ell + 1) \left[ \frac{D}{|C|} + \ln \frac{C}{\hat{C}} - 2 \right] ,
\label{eq:loglike}
\end{equation}
where $D = C_\ell^\text{TT} \hat{C}_\ell^\text{EE} + \hat{C}_\ell^\text{TT} C_\ell^\text{EE} - 2 C_\ell^\text{TE} \hat{C}_\ell^\text{TE}$, and we shift the likelihood such that 
\begin{equation}
\ln p \left(C=\hat{C}(\theta_0) | C \right) = 0,
\end{equation}
where we choose for our fiducial cosmology\footnote{For details see~\cite{Cole_2022}.}
\begin{equation}
\theta_0 = \left(\Omega_b h^2, \Omega_{\text{cdm}} h^2, 100 \theta_s, \ln(10^{10} A_s), n_s, \tau \right) = (0.0224, 0.12, 1.0411, 3.0753, 0.965, 0.054).
\end{equation}

\begin{figure} [h]
\centering
\includegraphics[scale = 0.7]{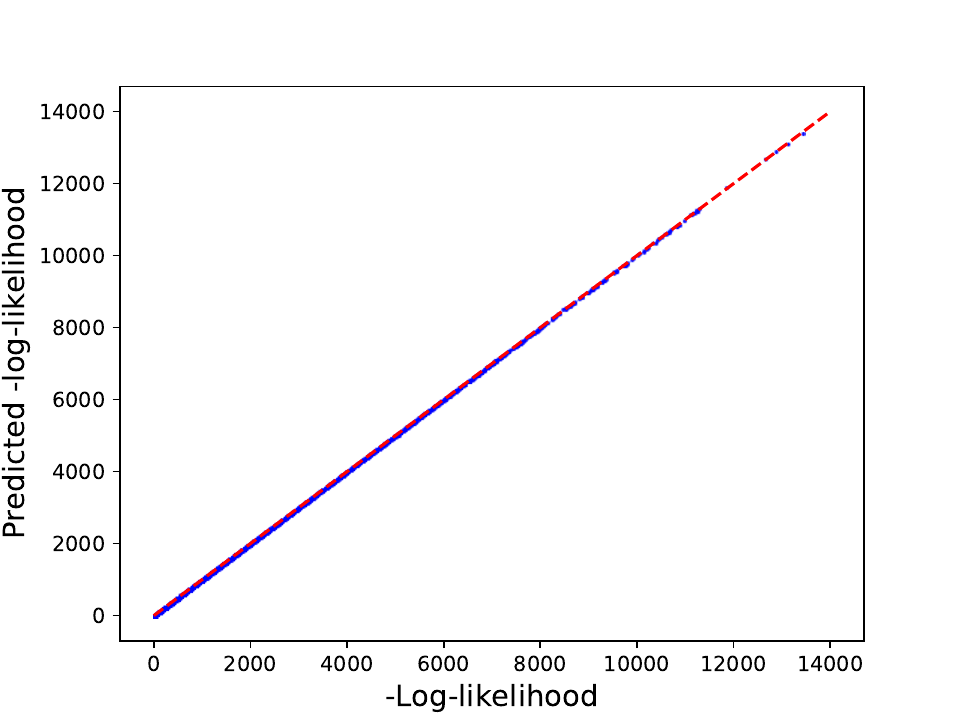}
\caption{Validation plot of the network, showing the true log likelihood on the x-axis and the predicted value on the y-axis.}
\label{fig:cmb_performance}
\end{figure}

The computational expense of evaluating the likelihood function represents a major challenge in inferring the parameters of cosmological models such as the six-dimensional $\Lambda$CDM model. To overcome this challenge, we have used a deep neural network to create a direct and efficient mapping from the cosmological parameter space to the log-likelihood defined in \cref{eq:loglike}. By training the network with many simulations, we bypass the need for computationally intensive direct simulations during the global fit itself. This strategy significantly speeds up the parameter estimation process, as has been demonstrated by previous work such as \code{CosmoPower}~\cite{SpurioMancini:2021ppk}

To this end, we have generated a large dataset of 1.2 million data points, each representing a unique configuration within the $\Lambda$CDM parameter space, based on a flat prior distribution as shown in \cref{tab:priors}. We used 80\% of this dataset for training the neural network and split the remaining 20\% evenly between validation and testing. Both the input and the output data were normalised to a Gaussian distribution. The architecture of our neural network is explicitly tailored to map $\Lambda$CDM parameters to their log-likelihood counterparts.

\begin{table}[ht]
\centering
\begin{tabular}{l c}
\hline
Parameter & Prior Range \\
\hline
$100 \theta_s$ & [1.03, 1.05] \\
$\Omega_b h^2$ & [0.020, 0.024] \\
$\Omega_c h^2$ & [0.10, 0.14] \\
$n_s$ & [0.9, 1.0] \\
$\ln(10^{10} A_s)$ & [2.98, 3.16] \\
$\tau$ & [0.010, 0.097] \\
\hline
\end{tabular}
\caption{Prior ranges for the Lambda CDM model parameters.}
\label{tab:priors}
\end{table}

The neural network model was developed with \code{TensorFlow} and our model consists of a sequential arrangement of six hidden layers, each with 100 neurons. These layers use the Scaled Exponential Linear Unit (SELU)~\cite{klambauer2017selfnormalizing}  activation function, which was chosen for its self-normalising properties that significantly improve training efficiency. The initial weights of these layers were set using the LeCun Normal initialiser to optimise the initial conditions for training the network. The output layer consists of a single neuron with a linear activation function for our regression task. Finally, the Mean Absolute Error was selected as the loss function. 

A multi-layered training optimisation strategy was used to refine the performance of the model. This included the use of the Adam optimiser~\cite{kingma2017adam} with an initial learning rate of 0.001. To further improve the accuracy of the model, the learning rate was halved with each complete run of the dataset. To minimise the risk of overfitting, training was also terminated prematurely if no improvement in validation loss was observed over a period of ten epochs.

The performance of our neural network on the test data set is visualised in \cref{fig:cmb_performance}, demonstrating the ability of the network to accurately predict the log-likelihood values. 
 
\begin{figure} [h]
\centering
\includegraphics[scale = 0.7]{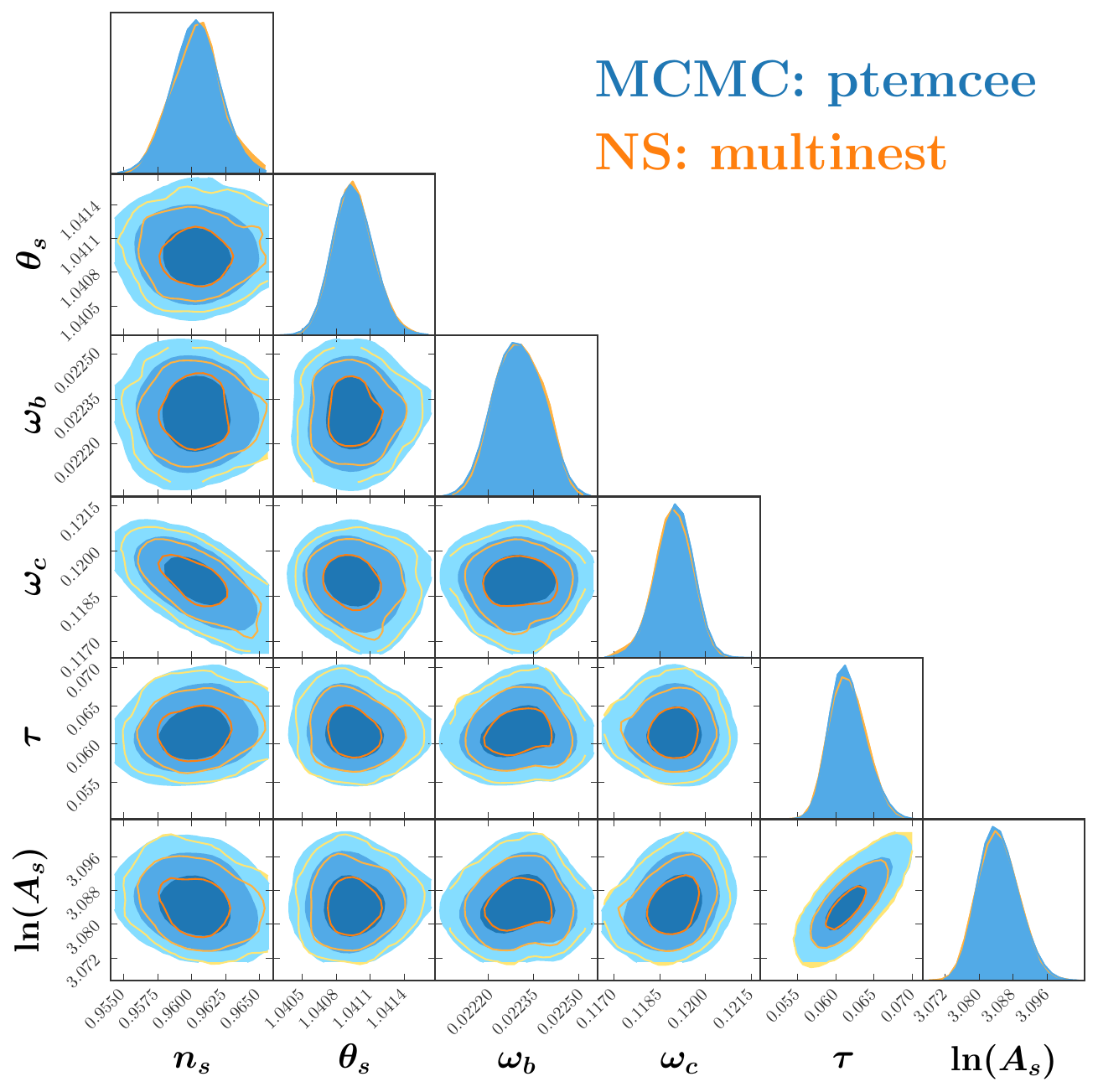}
\caption{Corner plot for the CMB Likelihood.}
\label{fig:cmb}
\end{figure}

A corner plot is shown in \cref{fig:cmb}, along with the posterior mean and error in \cref{tab:cmb_means} and summary metrics in \cref{tab:cmb_metrics}.

All algorithms converge effectively to the correct answer, as evidenced by the excellent agreement in the posterior mean, which is expected for a well-behaved unimodal function. Whilst neural network approximations can reliably reduce the computational load~\cite{Piras:2024dml} of historically expensive portions of the calculation, as we move further into the era of precision cosmology with more and more systematics to be modelled it will prove crucial to study which algorithms are the most efficient in terms of computational cost and performance. 

The most efficient algorithm was found to be \verb|Multinest|, which reached a satisfactory estimate of the posterior with only 8200 likelihood evaluations in total. The best MCMC algorithm was found to be \verb|ptemcee|, although \verb|zeus| and \verb|MCMC-diffusion| showed very similar performance. This matches the intuition that, for an unimodal posterior, the benefit of algorithms that allow for better mixing of modes is reduced. However, the evidence of this test shows that they still allow for a more efficient exploration of the posterior relative to more basic techniques. It is worth noting that the overall efficiency is relatively low for all the algorithms. 

Most importantly, this work shows that a CMB $\Lambda$CDM cosmological example is not a stringent test of a sampler's efficacy in comparison with the other examples considered in this paper, as the posterior is well-constrained in all parameters and approximately gaussian. This is not true for beyond $\Lambda$CDM models, or for non CMB data as is the case on both counts for the recent DESI analysis~\cite{DESI:2024mwx}.

\begin{table}[]
    \centering
    \begin{tabular}{ccccc}
    \toprule
    Algorithm & Diagnostic & \neff & \nlike & Efficiency\\
    \midrule
    MCMC: \verb|emcee| & \rhat = 1.01 & 265 & 62.0k & 0.00875\\
    \midrule
    \rowcolor{row-highlight}
    MCMC: \verb|ptemcee| & \rhat = 1.00 & 2.68k & 62.0k & 0.0432\\
    \midrule
    MCMC: \verb|zeus| & \rhat = 1.00 & 2.14k & 62.0k & 0.0345\\
    \midrule
    MCMC: \verb|MCMC-diffusion| & \rhat = 1.00 & 2.09k & 62.0k & 0.0337\\
    \midrule
    NS: \verb|Dynesty| & $p_{\rm insert}$ = 0.72 & 2k &  359k &  5.6$\times 10^{-3}$\\
    \midrule
    \rowcolor{row-highlight}
    NS: \verb|Multinest| & $p_{\rm insert}$ = 6.04$\times 10^{-7}$ & 861 & 8.2k & 0.1 \\
    \midrule
    NS: \verb|Polychord| & $p_{\rm insert}$ = 0.96 & 2k & 872k & 2.3$\times 10^{-3}$ \\
    \midrule
    NS: \verb|Nessai| & $p_{\rm insert}$ = 0.18 & 1.37k & 25.6k & 0.054\\ 
    \midrule
    NS: \verb|JaxNS| &  & 604 & 633k  & 9.5$\times 10^{-4}$ \\
    \midrule
    NS: \verb|Nautilus| &  & 10k & 82.7k & 0.12 \\
    \midrule
    NS: \verb|Ultranest| &  & 958 & 63k & 0.015 \\
    \midrule
    \bottomrule
    \end{tabular}
    \caption{Summary of algorithm performance for CMB likelihood}
    \label{tab:cmb_metrics}
\end{table}

%%%%%%%%%%%%%%%%%%%%%%%%%%%%%%%%%%%%%%
\subsection{Particle physics --- LHC}
%%%%%%%%%%%%%%%%%%%%%%%%%%%%%%%%%%%%%%
An additional example of the use of global fit techniques in particle astrophysics is the attempt to explore models of beyond-Standard Model particle physics. The Standard Model, which compromises a series of gauge field theories describing the matter and interactions in our universe, suffers from various theoretical challenges and is known to be incomplete. Beyond-Standard Model physics scenarios add extra fields and interactions, accompanied by various free parameters that can currently be constrained by the lack of evidence for new physics in particle colliders and other experiments. Accurately mapping the posterior of popular beyond-Standard Model theories is crucial for designing and optimising the future experiments that will uncover our next theory of particle physics.

To compare our Bayesian sampling techniques on a realistic example, we use the results of a previous global fit of the Minimal Supersymmetric Standard Model, performed by the GAMBIT collaboration~\cite{MSSM}. The 7 parameters of interest were the soft mass terms $M_2$, $m^2_{\tilde{f}}$, $m^2_{H_u}$, $m^2_{H_d}$ (defined at the scale $Q=1$~TeV), the trilinear couplings $A_{u_3}$, $A_{d_3}$ and $\tan\beta$ (defined at the scale $m_Z$). The fit also varied a number of nuisance parameters, including the top quark mass, the strong coupling constant, the nuclear matrix elements for WIMP-nucleon scattering and the local dark matter density, leading to 12 parameters in total. The fit included an extensive range of experimental constraints, including results from particle colliders, flavour physics, measurements of the cosmic microwave background and null observations from direct and indirect dark matter search experiments. Evaluating the likelihood of any particular parameter point was computationally expensive, and the total fit required extensive CPU hours. 

To ensure that we can compare our Bayesian sampling techniques with relatively minor computational effort, we interpolated the likelihood obtained in~\cite{MSSM} using a deep neural network, as originally proposed in~\cite{2012CoPhC.183..960B,Brooijmans:2020yij}. The network was used in a previous DarkMachines challenge, and details of the network architecture and validation can be found in Ref.~\cite{DarkMachinesHighDimensionalSamplingGroup:2021wkt}.

\begin{figure} [h]
\centering
\includegraphics[scale = 0.7]{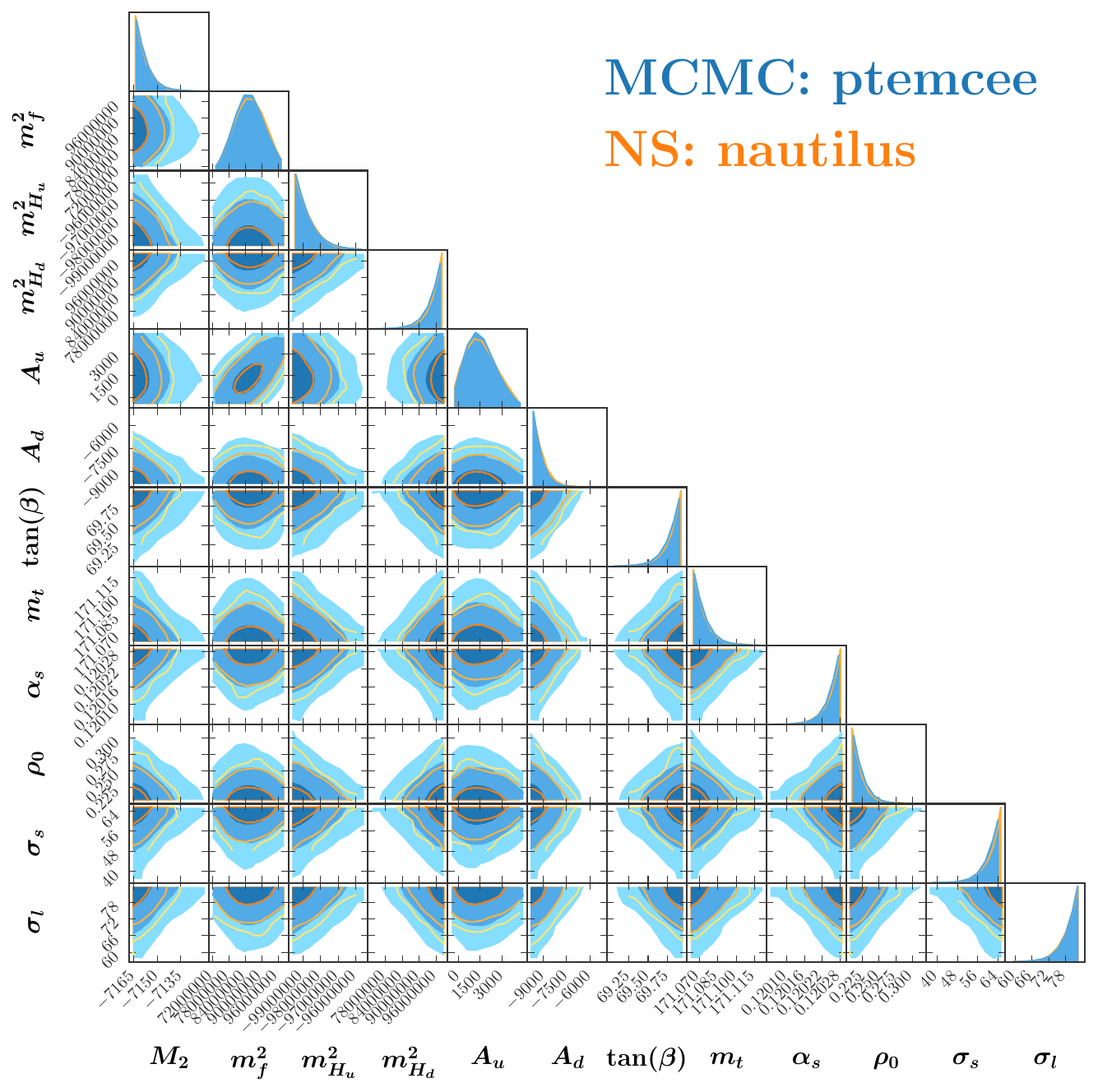}
\caption{Corner plot for the MSSM7 Likelihood.}
\label{fig:MSSM7}
\end{figure}

A corner plot is shown in \cref{fig:MSSM7}, along with the posterior mean and error in \cref{tab:mssm7_means} and summary metrics in \cref{tab:mssm7_metrics}. 

The results of this test show some distinct patterns of behaviour among the algorithms:

\begin{itemize}
    \item The posterior mean is broadly consistent between the \verb|ptemcee|, \verb|MCMC-diffusion|, \verb|Dynesty|, \verb|JaxNS|, \verb|Nautilus|, \verb|Ultranest| and \verb|PolyChord| algorithms, with the \verb|emcee| and \verb|zeus| values differing due to apparent lack of convergence. The likelihood surface for the MSSM7 is very complicated, with several disparate regions of high likelihood separated in the large volume of the parameter space, and this leads to a more complex posterior than that explored in any of the previous examples.
\item Even among the algorithms that broadly agree with each other,  the \verb|JaxNS| algorithm gives discrepant mean values for the parameters $m_f^2$, $m_{H_u}^2$, $m_{H_d}^2$, $A_u$, and $\tan (\beta)$, and has a large Wasserstein distance in parameters $m_f^2$ and $A_u$. It does not look wise to rely on the low-resolution results of \verb|JaxNS| in this case.
\item In terms of efficiency, \verb|Nautilus| is the most efficient algorithm, able to map the posterior satisfactorily with only 179,000 likelihood evaluations. For a likelihood that is as expensive as the original GAMBIT MSSM7 likelihood (which could take tens of minutes to hours per evaluation due to the need to simulate LHC proton-proton collisions), this efficiency is a crucial feature. The most efficient MCMC algorithm is \verb|ptemcee|, though again we see that the \verb|MCMC-diffusion| algorithm has a similar efficiency.
\end{itemize}

\begin{table}[]
    \centering
    \begin{tabular}{ccccc}
    \toprule
    Algorithm & Diagnostic & \neff & \nlike & Efficiency\\
    \midrule
    MCMC: \verb|emcee| & \rhat = 1.76 & 43 & 111k & 4.18 $\times 10^{-4}$\\
    \midrule
    \rowcolor{row-highlight}
    MCMC: \verb|ptemcee| & \rhat = 1.00 & 6.1k & 123k & 0.0495\\
    \midrule
    MCMC: \verb|zeus| & \rhat = 1.21 & 141 & 111k & 1.17$\times 10^{-3}$\\
    \midrule
    MCMC: \verb|MCMC-diffusion| & \rhat = 1.00 & 4.3k & 111k & 0.0387\\
    \midrule
    \verb|Dynesty| & $p_{\rm insert}$ = 0.15 & 3.5k &  969k & 3.7$\times 10^{-3}$\\
    \midrule
    \verb|JaxNS| &  & 1.9k & 3.9M  & 4.9$\times 10^{-4}$\\
    \midrule
    \rowcolor{row-highlight}
    \verb|Nautilus| &  & 10k & 179k & 0.0559 \\
    \midrule
    \verb|Ultranest| &  & 2.8k & 428k & 6.4$\times 10^{-3}$ \\
    \midrule
    \bottomrule
    \end{tabular}
    \caption{Summary of algorithm performance for MSSM7 likelihood}
    \label{tab:mssm7_metrics}
\end{table}

\section{Conclusions}
\label{sec:conclusions}

We have performed a detailed comparison of a variety of Bayesian sampling algorithms in an attempt
to determine to what extent each of these algorithms can accurately and efficiently describe complicated high-dimensional test functions and real-world physics likelihoods. We have approached this question by studying algorithmic performance out-of-the-box with minimal hyperparameter tuning and low computation time, so as to simulate the typical performance one might expect to quickly obtain as a physicist unfamiliar with the inner workings of each algorithm. On this basis we can reach a number of conclusions:

\begin{itemize}
    \item Modern MCMC tools such as \code{MCMC-diffusion} and \code{ptemcee} can tackle multi-modal problems, bringing MCMC efficiency to a realm previously only solvable with Nested Sampling. These modern MCMC algorithms tend to be more efficient than their NS counterparts across all the test functions studied here.
    \item The \code{emcee} and \code{zeus} algorithms are severely outclassed by the other algorithms present in this paper in terms of efficiency, and are rarely able to obtain reasonable results within the computational limits we have imposed.
    \item No NS algorithm has clearly emerged as the ``best'' in our results, rather certain algorithms appear more well suited to certain types of functions. For example, \code{Polychord} appears to be the most effective for a problem with differently weighted modes like the spike-slab, while \code{dynesty} appears to be the most effective for a heavily multimodal problem like the Eggbox.
    \item Subtlety exists even within NS when considering region-based sampling versus chain-based sampling with/without semi-independent mode tracking.
    \item The CMB+$\Lambda$CDM physics likelihood was accurately and efficiently sampled by all the algorithms tested here, so cosmologists dealing with similar unimodal problems need not be picky about which specific algorithm they choose among the options we have explored in this paper.
\end{itemize}

For both the
analytic functions and the two physics likelihoods explored in this paper, we have observed that many of the algorithms used here show promising performance. This certainly motivates their use in further
real-world physics applications, and we look forward to reading future examples
of their use.

\section{Acknowledgements}
The author(s) gratefully acknowledges the computer resources at Artemisa, funded by the European Union ERDF and Comunitat Valenciana as well as the technical support provided by the Instituto de Física Corpuscular, IFIC (CSIC-UV). R. RdA is supported by PID2020-113644GB-I00 from the Spanish Ministerio de Ciencia e Innovación and by the PROMETEO/2022/69 from the Spanish GVA. AF is supported by an NSFC Research Fund for International Young Scientists grant 11950410509. MJW is supported by the ARC Centre of Excellence for Dark Matter Particle Physics CE200100008, and CB by the ARC Discovery Projects DP210101636 and DP220100643. We also thank Johannes Lange for useful discussions on the \texttt{nautilus} algorithm.

\appendix

\section{Effective sample size}\label{sec:ess}

\subsection{MCMC}

The effective number of samples in an MCMC chain may be computed per parameter by taking into account the auto-correlation. We may define $\neff$ in this setting through the variance of a posterior mean,
\begin{equation}
\text{Var}(\bar x) = \frac{\text{Var}(x)}{\neff}
\end{equation}
This results in
\begin{equation}
\neff^\text{MCMC} = \frac{n}{1 + 2 \sum_{t=1}^\infty \rho(t)}
\end{equation}
where $\rho(t)$ is the auto-correlation at lag $t$. We again use \code{arviz}~\cite{arviz_2019} to compute this. This measure wouldn't be appropriate if one were particularly interested in samples from the tail of the distribution. To ensure that diagnostics and even $\neff$ itself may be computed reliably,~\cite{2019arXiv190308008V} recommend at least $400$ effective samples for any scalar quantity of interest. Since the auto-correlation differs per parameter, as a conservative overall effective sample size we take the minimum $\neff$ among the model parameters.

\subsection{Nested sampling}

In NS, we estimate the effective sample size using a standard rule of thumb~\cite{kish},
\begin{equation}\label{eq:neff_kish}
\neff^\text{Kish} = \frac{1}{\sum_{i=1}^N w_i^2}
\end{equation}
where $\sum w_{i=1}^N = 1$ and $w_i$ are the NS weights. This is a well-known heuristic that unfortunately lacks formal justification~\cite{https://doi.org/10.1111/insr.12500} and should be used as a `rough guide'~\cite{bda}. 
We thus furthermore consider an information-theoretic alternative,
\begin{equation}\label{eq:neff_information}
\neff^\text{Information} = e^{-\sum_{i=1}^N w_i \log w_i} 
\end{equation}
In both cases, equally weighted samples, $w_i = 1 / N$ result in $\neff = N$, and we expect $\neff \propto \nlive$. Viewing them as expectations, by Jensen's inequality,
\begin{equation}
\langle w \rangle \ge e^{\langle \ln w \rangle}
\end{equation}
such that the rule of thumb is more conservative, $\neff^\text{Kish} \le \neff^\text{Information}$. To simulate the weights $w_i$ we use \code{anesthetic}~\cite{anesthetic}.

Unlike for MCMC, these are overall estimates, not per parameter. Thus, finally we build an estimator more directly comparable to the MCMC case per parameter by considering
\begin{equation}\label{eq:neff_bootstrap}
\neff^\text{Bootstrap} = \frac{\text{Var}(x)}{\text{Var}(\bar x)}
\end{equation}
where $\text{Var}(x)$ is the posterior variance of a parameter, and $\text{Var}(\bar x)$ is the variance of our estimate of the posterior mean of the parameter. The latter takes into account two sources of variance: sampling of the compression factors, $t$, and sampling $x$ from around a likelihood contour~\cite{2018BayAn..13..873H} using a bootstrap procedure in \code{nestcheck}~\cite{higson2018nestcheck,higson2019diagnostic}.

There is a computational subtlety about estimating $\neff^\text{Kish}$ from NS results. If we write the weights as
\begin{equation}
w_i \propto L_i \Delta X_i = L_i X_i (1 - t_i)
\end{equation}
such that
\begin{equation}
\sum_{i=1}^N w_i^2 \propto \sum_{i=1}^N L_i^2 X_i^2 (1 - t_i)^2
\end{equation}
and use estimators for $t_i \simeq \langle t \rangle$ in this equation, we typically overestimate $\neff^\text{Kish}$ by about a factor of two as $\langle (1 - t)^2 \rangle / (1 - \langle t \rangle )^2 \simeq 2$. A similar error occurs if using $\ln t_i \simeq \langle \ln t \rangle$.

Finally, chain sampler NS implementations produce a chain of phantom points when generating a draw from the constrained prior. These phantom points may contribute to the posterior samples, but due to their correlations their contribution to the effective sample size is challenging to quantify.

\section{Posterior mean and standard deviation}\label{sec:post_mean}

\subsection{Eggbox}

\begin{table}[h!]
    \hspace*{-2.7cm}
    \tiny
    \centering
    \begin{tabular}{ccccccccccc}
    \toprule
    Algorithm & $\theta_1$ & $\theta_2$ & $\theta_3$ & $\theta_4$ & $\theta_5$ & $\theta_6$ & $\theta_7$ & $\theta_8$ & $\theta_9$ & $\theta_{10}$\\
    \midrule
    \rowcolor{row-highlight}
    True & 15.7 $\pm$ 9.3 & 15.7 $\pm$ 9.3 & 15.7 $\pm$ 9.3 & 15.7 $\pm$ 9.3 & 15.7 $\pm$ 9.3 & 15.7 $\pm$ 9.3 & 15.7 $\pm$ 9.3 & 15.7 $\pm$ 9.3 & 15.7 $\pm$ 9.3 & 15.7 $\pm$ 9.3\\
    \midrule
    MCMC: \verb|emcee| & 13.8 $\pm$ 7.8 & 15.1 $\pm$ 7.1 & 15.7 $\pm$ 8.6 & 15.3 $\pm$ 7.2 & 16.6 $\pm$ 8.5 & 15.1 $\pm$ 8.6 & 15.9 $\pm$ 7.7 & 16.5 $\pm$ 8.4 & 15.3 $\pm$ 7.8 & 16.1 $\pm$ 7.5\\
    \midrule
    MCMC: \verb|ptemcee| & 15.9 $\pm$ 7.4 & 16.1 $\pm$ 7.8 & 15.8 $\pm$ 7.9 & 15.9 $\pm$ 8.2 & 15.5 $\pm$ 7.5 & 15.7 $\pm$ 7.6 & 15.3 $\pm$ 8.0 & 16.6 $\pm$ 8.1 & 16.2 $\pm$ 7.7 & 15.2 $\pm$ 8.2\\
    \midrule
    MCMC: \verb|zeus| & 15.4 $\pm$ 9.2 & 15.5 $\pm$ 9.3 & 15.2 $\pm$ 9.4 & 15.7 $\pm$ 9.4 & 15.8 $\pm$ 9.2 & 15.8 $\pm$ 9.1 & 15.8 $\pm$ 9.2 & 16.5 $\pm$ 9.4 & 15.3 $\pm$ 9.2 & 16.1 $\pm$ 9.2 \\
    \midrule
    \rowcolor{row-highlight}
    MCMC: \verb|MCMC-diffusion| & 16.5 $\pm$ 9.8 & 15.4 $\pm$ 9.3 & 15.1 $\pm$ 9.2 & 14.4 $\pm$ 8.7 & 15.2 $\pm$ 9.5 & 15.5 $\pm$ 9.1 & 14.6 $\pm$ 8.9 & 16.0 $\pm$ 8.5 & 15.7 $\pm$ 8.3 & 16.6 $\pm$ 9.4\\
    \midrule
    \rowcolor{row-highlight}
    NS: \verb|dynesty| & 15.6 $\pm$ 9.3 & 15.7 $\pm$ 9.4 & 15.5 $\pm$ 9.3 & 15.9 $\pm$ 9.2 & 15.9 $\pm$ 9.3 & 15.7 $\pm$ 9.4 & 15.5 $\pm$ 9.3 & 15.5 $\pm$ 9.4 & 15.7 $\pm$ 9.4 & 16.2 $\pm$ 9.4 \\
    \bottomrule
    \end{tabular}
    \caption{Posterior mean for $10D$ Eggbox test function. The true mean for each parameter is $15.7 \pm 9.3$.}
    \label{tab:Eggbox_means}
\end{table}

\subsection{Rosenbrock}

\begin{table}[h!]
    \hspace*{-2.7cm}
    \tiny
    \centering
    \begin{tabular}{ccccccccccc}
    \toprule
    Algorithm & $\theta_1$ & $\theta_2$ & $\theta_3$ & $\theta_4$ & $\theta_5$ & $\theta_6$ & $\theta_7$ & $\theta_8$ & $\theta_9$ & $\theta_{10}$\\
    \midrule
    MCMC: \verb|emcee| & 0.66 $\pm$ 0.45 & 0.64 $\pm$ 0.32 & 0.52 $\pm$ 0.35 & 0.39 $\pm$ 0.35 & 0.28 $\pm$ 0.32 & 0.19 $\pm$ 0.27 & 0.12 $\pm$ 0.21 & 0.07 $\pm$ 0.15 & 0.03 $\pm$ 0.10 & 0.01 $\pm$ 0.08\\
    \midrule
    \rowcolor{row-highlight}
    MCMC: \verb|ptemcee| & 0.92 $\pm$ 0.33 & 0.94 $\pm$ 0.15 & 0.91 $\pm$ 0.19 & 0.86 $\pm$ 0.23 & 0.80 $\pm$ 0.28 & 0.72 $\pm$ 0.32 & 0.64 $\pm$ 0.36 & 0.55 $\pm$ 0.41 & 0.48 $\pm$ 0.49 & 0.46 $\pm$ 0.69\\
    \midrule
    MCMC: \verb|zeus| & 0.65 $\pm$ 0.68 & 0.88 $\pm$ 0.24 & 0.84 $\pm$ 0.28 & 0.78 $\pm$ 0.32 & 0.72 $\pm$ 0.32 & 0.64 $\pm$ 0.38 & 0.56 $\pm$ 0.40 & 0.48 $\pm$ 0.42 & 0.41 $\pm$ 0.48 & 0.39 $\pm$ 0.63\\
    \midrule
    MCMC: \verb|MCMC-diffusion| & 0.92 $\pm$ 0.31 & 0.93 $\pm$ 0.15 & 0.90 $\pm$ 0.20 & 0.86 $\pm$ 0.25 & 0.80 $\pm$ 0.28 & 0.73 $\pm$ 0.32 & 0.67 $\pm$ 0.34 & 0.56 $\pm$ 0.40 & 0.48 $\pm$ 0.49 & 0.50 $\pm$ 0.71\\
    \midrule
    \midrule
    \rowcolor{row-highlight}
    NS: \verb|dynesty| & 0.94 $\pm$ 0.22 & 0.93 $\pm$ 0.16 & 0.90 $\pm$ 0.20 & 0.86 $\pm$ 0.24 & 0.80 $\pm$ 0.29 & 0.72 $\pm$ 0.33 & 0.64 $\pm$ 0.37 & 0.55 $\pm$ 0.41 & 0.47 $\pm$ 0.48 & 0.45 $\pm$ 0.70 \\
    \midrule
    NS: \verb|multinest| & 0.91 $\pm$ 0.23 & 0.89 $\pm$ 0.20 & 0.83 $\pm$ 0.25 & 0.75 $\pm$ 0.30 & 0.66 $\pm$ 0.33 & 0.54 $\pm$ 0.34 & 0.42 $\pm$ 0.33 & 0.30 $\pm$ 0.30 & 0.18 $\pm$ 0.25 & 0.10 $\pm$ 0.20 \\
    \midrule
    NS: \verb|nessai| & 0.92 $\pm$ 0.19 & 0.89 $\pm$ 0.19 & 0.83 $\pm$ 0.24 & 0.76 $\pm$ 0.28 & 0.67 $\pm$ 0.31 & 0.54 $\pm$ 0.32 & 0.41 $\pm$ 0.31 & 0.27 $\pm$ 0.27 & 0.15 $\pm$ 0.21 & 0.07 $\pm$ 0.17 \\
    \midrule
    NS: \verb|JaxNS| & 0.93 $\pm$ 0.26 & 0.93 $\pm$ 0.14 & 0.89 $\pm$ 0.18 & 0.83 $\pm$ 0.22 & 0.73 $\pm$ 0.26 & 0.61 $\pm$ 0.29 & 0.46 $\pm$ 0.29 & 0.30 $\pm$ 0.26 & 0.16 $\pm$ 0.20 & 0.06 $\pm$ 0.15 \\
    \midrule
    NS: \verb|nautilus| & 0.93 $\pm$ 0.26 & 0.93 $\pm$ 0.16 & 0.90 $\pm$ 0.20 & 0.85 $\pm$ 0.24 & 0.79 $\pm$ 0.29 & 0.71 $\pm$ 0.33 & 0.62 $\pm$ 0.36 & 0.51 $\pm$ 0.38 & 0.21 $\pm$ 0.41 & 0.11 $\pm$ 0.50 \\
    \bottomrule
    \end{tabular}
    \caption{Posterior mean for $10D$ Rosenbrock test function.}
    \label{tab:Rosenbrock_means}
\end{table}

\newpage

\subsection{Rastrigin}

\begin{table}[h!]
    \hspace*{-2.7cm}
    \tiny
    \centering
    \begin{tabular}{ccccccccccc}
    \toprule
    Algorithm & $\theta_1$ & $\theta_2$ & $\theta_3$ & $\theta_4$ & $\theta_5$ & $\theta_6$ & $\theta_7$ & $\theta_8$ & $\theta_9$ & $\theta_{10}$\\
    \midrule
    \rowcolor{row-highlight}
    True & 0.00 $\pm$ 0.69 & 0.00 $\pm$ 0.69 & 0.00 $\pm$ 0.69 & 0.00 $\pm$ 0.69 & 0.00 $\pm$ 0.69 & 0.00 $\pm$ 0.69 & 0.00 $\pm$ 0.69 & 0.00 $\pm$ 0.69 & 0.00 $\pm$ 0.69 & 0.00 $\pm$ 0.69\\
    \midrule
    MCMC: \verb|emcee| & 0.27 $\pm$ 0.93 & -0.37 $\pm$ 0.92 & -0.26 $\pm$ 0.85 & -0.04 $\pm$ 0.99 & 0.37 $\pm$ 0.87 & -0.09 $\pm$ 1.23 & -0.53 $\pm$ 0.67 & 0.10 $\pm$ 0.65 & 0.22 $\pm$ 1.03 & 0.35 $\pm$ 1.10\\
    \midrule
    \rowcolor{row-highlight}
    MCMC: \verb|ptemcee| & -0.03 $\pm$ 0.69 & 0.01 $\pm$ 0.74 & 0.03 $\pm$ 0.70 & 0.02 $\pm$ 0.69 & 0.00 $\pm$ 0.70 & -0.02 $\pm$ 0.70 & -0.05 $\pm$ 0.70 & 0.01 $\pm$ 0.69 & 0.03 $\pm$ 0.73 & 0.08 $\pm$ 0.67\\
    \midrule
    MCMC: \verb|zeus| & 0.50 $\pm$ 2.39 & -0.36 $\pm$ 2.65 & 0.03 $\pm$ 2.81 & -0.36 $\pm$ 2.65 & 0.27 $\pm$ 2.67 & -0.43 $\pm$ 3.00 & -1.38 $\pm$ 2.70 & -0.90 $\pm$ 2.97 & -0.43 $\pm$ 2.60 & 0.57 $\pm$ 2.95\\
    \midrule
    MCMC: \verb|MCMC-diffusion| & -0.09 $\pm$ 0.82 & 0.24 $\pm$ 0.99 & 0.05 $\pm$ 0.72 & 0.14 $\pm$ 0.55 & -0.21 $\pm$ 0.69 & 0.02 $\pm$ 0.71 & 0.17 $\pm$ 0.84 & -0.38 $\pm$ 1.05 & -0.06 $\pm$ 0.78 & 0.09 $\pm$ 0.72\\
    \midrule
    \rowcolor{row-highlight}
    NS: \verb|JaxNS| & 0.13 $\pm$ 0.71 & 0.09 $\pm$ 0.76 & -0.27 $\pm$ 0.89 & -0.07 $\pm$ 0.71 & -0.05 $\pm$ 0.72 & -0.16 $\pm$ 0.64 & 0.16 $\pm$ 0.74 & -0.04 $\pm$ 0.56 & -0.31 $\pm$ 0.71 & -0.12 $\pm$ 0.76\\
    \bottomrule
    \end{tabular}
    \caption{Posterior mean for $10D$ Rastrigin test function. The true mean for each parameter is 0. }
    \label{tab:Rastrigin_means}
\end{table}

\subsection{Spike-slab function}

\begin{table}[h!]
    \hspace*{-2.5cm}
    \tiny
    \centering
    \begin{tabular}{ccccccccccc}
    \toprule
    Algorithm & $\theta_1$ & $\theta_2$ & $\theta_3$ & $\theta_4$ & $\theta_5$ & $\theta_6$ & $\theta_7$ & $\theta_8$ & $\theta_9$ & $\theta_{10}$\\
    \midrule
    \rowcolor{row-highlight}
    True & 4.25 $\pm$ 1.84 & 4.25 $\pm$ 1.84 & 0.00 $\pm$ 0.57 & 0.00 $\pm$ 0.57 & 0.00 $\pm$ 0.57 & 0.00 $\pm$ 0.57 & 0.00 $\pm$ 0.57 & 0.00 $\pm$ 0.57 & 0.00 $\pm$ 0.57 & 0.00 $\pm$ 0.57\\
    \midrule
    MCMC: \verb|emcee| & 2.60 $\pm$ 1.09 & 2.62 $\pm$ 1.08 & 0.01 $\pm$ 0.88 & -0.00 $\pm$ 0.88 & 0.04 $\pm$ 0.88 & 0.01 $\pm$ 0.89 & 0.00 $\pm$ 0.87 & -0.01 $\pm$ 0.88 & -0.01 $\pm$ 0.87 & 0.00 $\pm$ 0.88\\
    \midrule
    MCMC: \verb|ptemcee| & 2.52 $\pm$ 0.94 & 2.53 $\pm$ 0.93 & 0.01 $\pm$ 0.88 & -0.01 $\pm$ 0.89 & -0.03 $\pm$ 0.89 & 0.01 $\pm$ 0.88 & 0.00 $\pm$ 0.89 & 0.01 $\pm$ 0.89 & -0.02 $\pm$ 0.89 & 0.02 $\pm$ 0.89\\
    \midrule
    MCMC: \verb|zeus| & 3.20 $\pm$ 1.62 & 3.20 $\pm$ 1.62 & 0.01 $\pm$ 0.82 & 0.01 $\pm$ 0.82 & 0.0 $\pm$ 0.82 & -0.01 $\pm$ 0.82 & 0.00 $\pm$ 0.82 & 0.00 $\pm$ 0.83 & 0.01 $\pm$ 0.82 & -0.01 $\pm$ 0.82\\
    \midrule
    \rowcolor{row-highlight}
    MCMC: \verb|MCMC-diffusion| & 4.38 $\pm$ 1.81 & 4.45 $\pm$ 1.92 & -0.02 $\pm$ 0.72 & -0.02 $\pm$ 0.72 & 0.03 $\pm$ 0.73 & -0.01 $\pm$ 0.73 & -0.00 $\pm$ 0.74 & 0.02 $\pm$ 0.73 & 0.00 $\pm$ 0.74 & 0.00 $\pm$ 0.74\\
    \midrule
    NS: \verb|Polychord| & 4.80 $\pm$ 1.74 & 4.80 $\pm$ 1.76 & 0.00 $\pm$ 0.56 & 0.00 $\pm$ 0.56 & 0.00 $\pm$ 0.56 & -0.01 $\pm$ 0.56 & 0.00 $\pm$ 0.57 & 0.01 $\pm$ 0.57 & -0.02 $\pm$ 0.57 & 0.01 $\pm$ 0.57\\
    \midrule
    \rowcolor{row-highlight}
    NS: \verb|JaxNS| & 5.11 $\pm$ 1.60 & 5.11 $\pm$ 1.61 & 0.00 $\pm$ 0.50 & 0.00 $\pm$ 0.51 & 0.01 $\pm$ 0.50 & -0.01 $\pm$ 0.51 & 0.00 $\pm$ 0.51 & 0.00 $\pm$ 0.50 & 0.00 $\pm$ 0.51 & 0.00 $\pm$ 0.51\\
    \bottomrule
    \end{tabular}
    \caption{Posterior mean and error for $10D$ spike-slab test function.}
    \label{tab:spikeslab_means}
\end{table}

\newpage

\begin{figure}
    \centering
    \includegraphics[width=\textwidth]{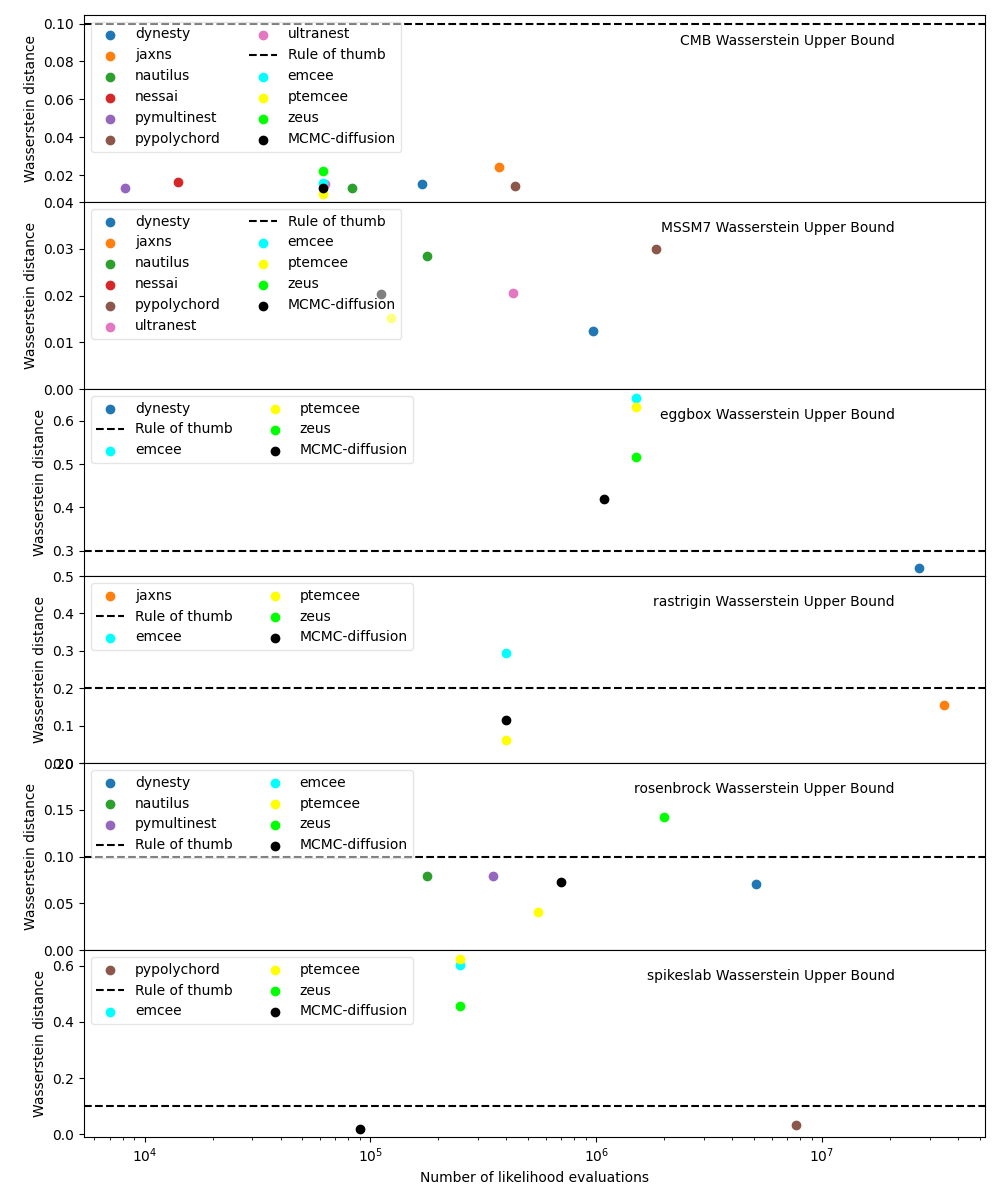}
    \caption{Upper bound on Wasserstein distance from reference posterior for each problem. The black dashed line is the assumed rule of thumb for convergence for each problem in the unit-cube parameterised domain.}
    \label{fig:wasserstein_upper}
\end{figure}

\begin{table}[h!]
    \hspace*{-2.5cm}
    \tiny
    \centering
    \begin{tabular}{ccccccc}
    \toprule
    Algorithm & $n_s$ & $\theta_s$ & $\omega_b$ & $\omega_c$ & $\tau$ & $\ln(A_s)$ \\\\
    \midrule
    MCMC:emcee & 0.0066 & 0.0031 & 0.0030 & 0.00081 & 0.0011 & 0.0011 \\
    \midrule
    MCMC:ptemcee & 0.0046 & 0.0011 & 0.0010 & 0.00047 & 0.0016 & 0.0017 \\
    \midrule
    MCMC:zeus & 0.0062 & 0.0036 & 0.0029 & 0.0020 & 0.0043 & 0.0035 \\
    \midrule
    MCMC:MCMC-diffusion & 0.0063 & 0.0019 & 0.0015 & 0.00064 & 0.0015 & 0.0017 \\
    \midrule
    dynesty & 0.0084 & 0.0014 & 0.0009 & 0.0007 & 0.0020 & 0.0022 \\
    \midrule
    jaxns & 0.0136 & 0.0045 & 0.0013 & 0.0005 & 0.0022 & 0.0023 \\
    \midrule
    nautilus & 0.0048 & 0.0013 & 0.0012 & 0.0008 & 0.0030 & 0.0022 \\
    \midrule
    nessai & 0.0088 & 0.0013 & 0.0016 & 0.0008 & 0.0017 & 0.0024 \\
    \midrule
    pymultinest & 0.0050 & 0.0011 & 0.0016 & 0.0006 & 0.0026 & 0.0025 \\
    \midrule
    pypolychord & 0.0066 & 0.0021 & 0.0017 & 0.0006 & 0.0018 & 0.0018 \\
    \midrule
    ultranest & 0.0053 & 0.0024 & 0.0017 & 0.0005 & 0.0026 & 0.0028 \\
    \bottomrule
    \end{tabular}
    \caption{Marginal Wasserstein-1 distances for each parameter from reference posterior for CMB for cheapest converging NS runs. An `--' indicates none of the completed runs converged.}
    \label{tab:CMB_wasserstein}
\end{table}

\begin{table}[h!]
    \hspace*{-2.5cm}
    \tiny
    \centering
    \begin{tabular}{ccccccccccccc}
    \toprule
    Algorithm & $M_2$ & $m_f^2$ & $m_{H_u}^2$ & $m_{H_d}^2$ & $A_u$ & $A_d$  & $\tan (\beta)$ & $m_t$ & $\alpha_s$ & $\rho_0$ & $\sigma_s$ & $\sigma_l$\\\\
    \midrule
    MCMC:emcee & 0.62 & 0.25 & 0.81 & 0.64 & 0.23 & 0.55 & 0.57 & 0.45 & 0.49 & 0.32 & 0.44 & 0.41 \\
    \midrule
    MCMC:ptemcee & 0.0001 & 0.0013 & 0.0006 & 0.0019 & 0.0008 & 0.0003 & 0.0002 & 0.0002 & 0.0006 & 0.0023 & 0.0038 & 0.0031 \\
    \midrule
    MCMC:zeus & 0.81 & 0.33 & 0.85 & 0.77 & 0.32 & 0.47 & 0.76 & 0.41 & 0.44 & 0.40 & 0.42 & 0.43  \\
    \midrule
    MCMC:MCMC-diffusion & 0.0000 & 0.0024 & 0.0008 & 0.0026 & 0.0010 & 0.0007 & 0.0001 & 0.0001 & 0.0008 & 0.0023 & 0.0063 & 0.0029 \\
    \midrule
    dynesty & 0.0000 & 0.0012 & 0.0005 & 0.0016 & 0.0011 & 0.0009 & 0.0002 & 0.0002 & 0.0004 & 0.0023 & 0.0027 & 0.0013 \\
    \midrule
    jaxns & 0.0000 & 0.0096 & 0.0003 & 0.0027 & 0.0095 & 0.0030 & 0.0000 & 0.0001 & 0.0003 & 0.0004 & 0.0123 & 0.0336 \\
    \midrule
    nautilus & 0.0000 & 0.0039 & 0.0008 & 0.0011 & 0.0044 & 0.0024 & 0.0003 & 0.0002 & 0.0012 & 0.0028 & 0.0071 & 0.0044 \\
    \midrule
    nessai & 0.0001 & 0.0081 & 0.0007 & 0.0023 & 0.0050 & 0.0053 & 0.0005 & 0.0005 & 0.0014 & 0.0021 & 0.0178 & 0.0107 \\
    \midrule
    pymultinest & -- & -- & -- & -- & -- & -- & -- & -- & -- & -- & -- & -- \\
    \midrule
    pypolychord & 0.0001 & 0.0022 & 0.0011 & 0.0021 & 0.0027 & 0.0027 & 0.0004 & 0.0003 & 0.0015 & 0.0021 & 0.0087 & 0.0061 \\
    \midrule
    ultranest & 0.0001 & 0.0019 & 0.0007 & 0.0019 & 0.0016 & 0.0017 & 0.0002 & 0.0001 & 0.0008 & 0.0024 & 0.0059 & 0.0033 \\
    \bottomrule
    \end{tabular}
    \caption{Marginal Wasserstein-1 distances for each parameter from reference posterior for MSSM7 for cheapest converging NS runs. An `--' indicates none of the completed runs converged.}
    \label{tab:MSSM7_wasserstein}
\end{table}

\begin{table}[h!]
    \hspace*{-2.5cm}
    \tiny
    \centering
    \begin{tabular}{ccccccccccc}
    \toprule
    Algorithm & $\theta_{1}$ & $\theta_{2}$ & $\theta_{3}$ & $\theta_{4}$ & $\theta_{5}$ & $\theta_{6}$ & $\theta_{7}$ & $\theta_{8}$ & $\theta_{9}$ & $\theta_{10}$ \\\\
    \midrule
    MCMC:emcee & 0.092 & 0.087 & 0.048 & 0.087 & 0.043 & 0.048 & 0.065 & 0.046 & 0.057 & 0.082 \\
    \midrule
    MCMC:ptemcee & 0.086 & 0.062 & 0.062 & 0.055 & 0.073 & 0.050 & 0.070 & 0.059 & 0.060 & 0.053 \\
    \midrule
    MCMC:zeus & 0.066 & 0.028 & 0.11 & 0.040 & 0.042 & 0.044 & 0.078 & 0.032 & 0.030 & 0.050 \\
    \midrule
    MCMC:MCMC-diffusion & 0.029 & 0.019 & 0.057 & 0.047 & 0.036 & 0.022 & 0.075 & 0.040 & 0.045 & 0.050 \\
    \midrule
    dynesty & 0.0272 & 0.0175 & 0.0357 & 0.0354 & 0.0208 & 0.0189 & 0.0472 & 0.0134 & 0.0158 & 0.0292 \\
    \midrule
    jaxns & -- & -- & -- & -- & -- & -- & -- & -- & -- & -- \\
    \midrule
    nautilus & -- & -- & -- & -- & -- & -- & -- & -- & -- & -- \\
    \midrule
    nessai & -- & -- & -- & -- & -- & -- & -- & -- & -- & -- \\
    \midrule
    pymultinest & -- & -- & -- & -- & -- & -- & -- & -- & -- & -- \\
    \midrule
    pypolychord & -- & -- & -- & -- & -- & -- & -- & -- & -- & -- \\
    \midrule
    ultranest & -- & -- & -- & -- & -- & -- & -- & -- & -- & -- \\
    \bottomrule
    \end{tabular}
    \caption{Marginal Wasserstein-1 distances for each parameter from reference posterior for Eggbox for cheapest converging NS runs. An `--' indicates none of the completed runs converged.}
    \label{tab:eggbox_wasserstein}
\end{table}

\begin{table}[h!]
    \hspace*{-2.5cm}
    \tiny
    \centering
    \begin{tabular}{ccccccccccc}
    \toprule
    Algorithm & $\theta_{1}$ & $\theta_{2}$ & $\theta_{3}$ & $\theta_{4}$ & $\theta_{5}$ & $\theta_{6}$ & $\theta_{7}$ & $\theta_{8}$ & $\theta_{9}$ & $\theta_{10}$ \\\\
    \midrule
    MCMC:emcee & 0.037 & 0.027 & 0.027 & 0.026 & 0.027 & 0.026 & 0.036 & 0.026 & 0.034 & 0.030 \\
    \midrule
    MCMC:ptemcee & 0.0090 & 0.0062 & 0.0071 & 0.011 & 0.0052 & 0.0048 & 0.0040 & 0.0024 & 0.0076 & 0.0040 \\
    \midrule
    MCMC:zeus & 0.20 & 0.020 & 0.20 & 0.20 & 0.19 & 0.19 & 0.19 & 0.19 & 0.20 & 0.20 \\
    \midrule
    MCMC:MCMC-diffusion & 0.013 & 0.0060 & 0.010 & 0.0072 & 0.015 & 0.011 & 0.0068 & 0.010 & 0.012 & 0.013 \\
    \midrule
dynesty & -- & -- & -- & -- & -- & -- & -- & -- & -- & -- \\
    \midrule
    jaxns & 0.0193 & 0.0147 & 0.0391 & 0.0070 & 0.0130 & 0.0072 & 0.0220 & 0.0132 & 0.0151 & 0.0045 \\
    \midrule
    nautilus & -- & -- & -- & -- & -- & -- & -- & -- & -- & -- \\
    \midrule
    nessai & -- & -- & -- & -- & -- & -- & -- & -- & -- & -- \\
    \midrule
    pymultinest & -- & -- & -- & -- & -- & -- & -- & -- & -- & -- \\
    \midrule
    pypolychord & -- & -- & -- & -- & -- & -- & -- & -- & -- & -- \\
    \midrule
    ultranest & -- & -- & -- & -- & -- & -- & -- & -- & -- & -- \\
    \bottomrule
    \end{tabular}
    \caption{Marginal Wasserstein-1 distances for each parameter from reference posterior for rastrigin for cheapest converging NS runs. An `--' indicates none of the completed runs converged.}
    \label{tab:rastrigin_wasserstein}
\end{table}

\begin{table}[h!]
    \hspace*{-2.5cm}
    \tiny
    \centering
    \begin{tabular}{ccccccccccc}
    \toprule
    Algorithm & $\theta_{1}$ & $\theta_{2}$ & $\theta_{3}$ & $\theta_{4}$ & $\theta_{5}$ & $\theta_{6}$ & $\theta_{7}$ & $\theta_{8}$ & $\theta_{9}$ & $\theta_{10}$ \\\\
    \midrule
    MCMC:emcee & 0.028 & 0.031 & 0.041 & 0.049 & 0.055 & 0.058 & 0.058 & 0.055 & 0.052 & 0.059 \\
    \midrule
    MCMC:ptemcee & 0.0025 & 0.0010 & 0.0014 & 0.0018 & 0.0025 & 0.0032 & 0.0041 & 0.0051 & 0.0070 & 0.012 \\
    \midrule
    MCMC:zeus & 0.028 & 0.0066 & 0.0084 & 0.010 & 0.011 & 0.012 & 0.013 & 0.014 & 0.015 & 0.021 \\
    \midrule
    MCMC:MCMC-diffusion & 0.0032 & 0.0012 & 0.0021 & 0.0036 & 0.0044 & 0.0050 & 0.0069 & 0.0098 & 0.016 & 0.021 \\
    \midrule
    dynesty & 0.0011 & 0.0018 & 0.0026 & 0.0038 & 0.0054 & 0.0068 & 0.0082 & 0.0095 & 0.012 & 0.019 \\
    \midrule
    jaxns & -- & -- & -- & -- & -- & -- & -- & -- & -- & -- \\
    \midrule
    nautilus & 0.0015 & 0.0018 & 0.0026 & 0.0034 & 0.0046 & 0.0063 & 0.0080 & 0.0104 & 0.0147 & 0.0258 \\
    \midrule
    nessai & -- & -- & -- & -- & -- & -- & -- & -- & -- & -- \\
    \midrule
    pymultinest & 0.0021 & 0.0022 & 0.0031 & 0.0041 & 0.0056 & 0.0070 & 0.0086 & 0.011 & 0.014 & 0.023 \\
    \midrule
    pypolychord & -- & -- & -- & -- & -- & -- & -- & -- & -- & -- \\
    \midrule
    ultranest & -- & -- & -- & -- & -- & -- & -- & -- & -- & -- \\
    \bottomrule
    \end{tabular}
    \caption{Marginal Wasserstein-1 distances for each parameter from reference posterior for rosenbrock for cheapest converging NS runs. An `--' indicates none of the completed runs converged.}
    \label{tab:rosenbrock_wasserstein}
\end{table}

\begin{table}[h!]
    \hspace*{-2.5cm}
    \tiny
    \centering
    \begin{tabular}{ccccccccccc}
    \toprule
    Algorithm & $\theta_{1}$ & $\theta_{2}$ & $\theta_{3}$ & $\theta_{4}$ & $\theta_{5}$ & $\theta_{6}$ & $\theta_{7}$ & $\theta_{8}$ & $\theta_{9}$ & $\theta_{10}$ \\\\
    \midrule
    MCMC:emcee & 0.19 & 0.19 & 0.028 & 0.028 & 0.028 & 0.028 & 0.028 & 0.028 & 0.027 & 0.028 \\
    \midrule
    MCMC:ptemcee & 0.20 & 0.20 & 0.028 & 0.029 & 0.029 & 0.029 & 0.029 & 0.029 & 0.029 & 0.029 \\
    \midrule
    MCMC:zeus & 0.14 & 0.14 & 0.022 & 0.021 & 0.022 & 0.021 & 0.021 & 0.021 & 0.022 & 0.022 \\
    \midrule
    MCMC:MCMC-diffusion & 0.0046 & 0.0041 & 0.0013 & 0.0013 & 0.0013 & 0.0016 & 0.0014 & 0.0016 & 0.0013 & 0.0014 \\
    \midrule
    dynesty & -- & -- & -- & -- & -- & -- & -- & -- & -- & -- \\
    \midrule
    jaxns & -- & -- & -- & -- & -- & -- & -- & -- & -- & -- \\
    \midrule
    nautilus & -- & -- & -- & -- & -- & -- & -- & -- & -- & -- \\
    \midrule
    nessai & -- & -- & -- & -- & -- & -- & -- & -- & -- & -- \\
    \midrule
    pymultinest & -- & -- & -- & -- & -- & -- & -- & -- & -- & -- \\
    \midrule
    pypolychord & 0.0072 & 0.0077 & 0.0021 & 0.0018 & 0.0020 & 0.0023 & 0.0023 & 0.0023 & 0.0021 & 0.0023 \\
    \midrule
    ultranest & -- & -- & -- & -- & -- & -- & -- & -- & -- & -- \\
    \bottomrule
    \end{tabular}
    \caption{Marginal Wasserstein-1 distances for each parameter from reference posterior for spikeslab for cheapest converging NS runs. An `--' indicates none of the completed runs converged.}
    \label{tab:spikeslab_wasserstein}
\end{table}

\newpage

\subsection{Cosmology-Planck}

\begin{table}[h!]
    \tiny
    \centering
    \begin{tabular}{ccccccc}
    \toprule
    Algorithm & $n_s$ & $\theta_s$ & $\omega_b$ & $\omega_c$ & $\tau$ & $\ln(A_s)$\\
    \midrule
    MCMC: \verb|emcee| & 0.960 $\pm$ 0.001 & 1.04 $\pm$ 0.00 & 0.0222 $\pm$ 0.0001 & 0.119 $\pm$ 0.001 & 0.0616 $\pm$ 0.0021 & 3.08 $\pm$ 0.01 \\
    \midrule
    \rowcolor{row-highlight}
    MCMC: \verb|ptemcee| & 0.960 $\pm$ 0.001 & 1.04 $\pm$ 0.00 & 0.0223 $\pm$ 0.0001 & 0.119 $\pm$ 0.001 & 0.0616 $\pm$ 0.0021 & 3.08 $\pm$ 0.01 \\
    \midrule
    MCMC: \verb|zeus| & 0.960 $\pm$ 0.001 & 1.04 $\pm$ 0.00 & 0.0223 $\pm$ 0.0001 & 0.119 $\pm$ 0.001 & 0.0617 $\pm$ 0.0021 & 3.08 $\pm$ 0.01 \\
    \midrule
    MCMC: \verb|MCMC-diffusion| & 0.960 $\pm$ 0.001 & 1.04 $\pm$ 0.00 & 0.0223 $\pm$ 0.0001 & 0.119 $\pm$ 0.001 & 0.0618 $\pm$ 0.0021 & 3.09 $\pm$ 0.01 \\
    \midrule
    NS: \verb|Dynesty| & 0.960 $\pm$ 0.002 & 1.04 $\pm$ 0.00 & 0.0223 $\pm$ 0.0001 & 0.119 $\pm$ 0.001 & 0.0617 $\pm$ 0.0021 & 3.08 $\pm$ 0.01 \\
    \midrule
    \rowcolor{row-highlight}
    NS: \verb|Multinest| & 0.960 $\pm$ 0.002 & 1.04 $\pm$ 0.00 & 0.0223 $\pm$ 0.0001 & 0.119 $\pm$ 0.001 & 0.0617 $\pm$ 0.0020 & 3.08 $\pm$ 0.00 \\
    \midrule
    NS: \verb|Polychord| & 0.960 $\pm$ 0.001 & 1.04 $\pm$ 0.00 & 0.0223 $\pm$ 0.0001 & 0.119 $\pm$ 0.001 & 0.0615 $\pm$ 0.0021 & 3.08 $\pm$ 0.01 \\
    \midrule
    NS: \verb|Nessai| & 0.960 $\pm$ 0.002 & 1.04 $\pm$ 0.00 & 0.0223 $\pm$ 0.0001 & 0.119 $\pm$ 0.001 & 0.0615 $\pm$ 0.0020 & 3.08 $\pm$ 0.00 \\
    \midrule
    NS: \verb|JaxNS| & 0.960 $\pm$ 0.001 & 1.04 $\pm$ 0.00 & 0.0223 $\pm$ 0.0001 & 0.119 $\pm$ 0.001 & 0.0616 $\pm$ 0.0021 & 3.08 $\pm$ 0.00 \\
    \midrule
    NS: \verb|Nautilus| & 0.960 $\pm$ 0.02 & 1.04 $\pm$ 0.00 & 0.0223 $\pm$ 0.0001 & 0.119 $\pm$ 0.001 & 0.0617 $\pm$ 0.0022 & 3.08 $\pm$ 0.00 \\
    \midrule
    NS: \verb|Ultranest| & 0.960 $\pm$ 0.002 & 1.04 $\pm$ 0.00 & 0.0223 $\pm$ 0.0001 & 0.119 $\pm$ 0.001 & 0.0615 $\pm$ 0.0021 & 3.08 $\pm$ 0.00 \\
    \bottomrule
    \end{tabular}
    \caption{Posterior mean and error for the CMB likelihood.}
    \label{tab:cmb_means}
\end{table}

\subsection{Particle physics - LHC}

\begin{landscape}
\newpage
\begin{table}[p]
    % \hspace*{-3.8cm}
    % \tiny
    \fontsize{4}{4}\selectfont
    \centering
    \begin{tabular}{ccccccccccccc}
    \toprule
    Algorithm & $M_2$ & $m_f^2$ & $m_{H_u}^2$ & $m_{H_d}^2$ & $A_u$ & $A_d$  & $\tan (\beta)$ & $m_t$ & $\alpha_s$ & $\rho_0$ & $\sigma_s$ & $\sigma_l$\\
    \midrule
    MCMC: \verb|emcee| & 1730 $\pm$ 3800 & 6.08 $\times 10^7  \pm$ 2.67 $\times 10^7$ & -1.80 $\times 10^7  \pm$ 2.36 $\times 10^7$ & -1.05 $\times 10^7  \pm$ 5.02 $\times 10^7$ & 1170 $\pm$ 6190 & 1500 $\pm$ 5240 & 31.5 $\pm$ 18.4 & 173 $\pm$ 1.13 & 0.119 $\pm$ 0.001 & 0.402 $\pm$ 0.143 & 42.3 $\pm$ 12.0 & 59.0 $\pm$ 16.1\\
    \midrule
    \rowcolor{row-highlight}
    MCMC: \verb|ptemcee| & -7160 $\pm$ 10 & 8.56 $\times 10^7  \pm$ 0.49 $\times 10^7$ & -9.91 $\times 10^7  \pm$ 0.08 $\times 10^7$ & 9.66 $\times 10^7  \pm$ 0.34 $\times 10^7$ & 1510 $\pm$ 970 & -9510 $\pm$ 500 & 69.9 $\pm$ 0.1 & 171 $\pm$ 0.01 & 0.120 $\pm$ 0.001 & 0.212 $\pm$ 0.012 & 63.1 $\pm$ 3.8 & 81.6 $\pm$ 3.73\\
    \midrule
    MCMC: \verb|zeus| & 4390 $\pm$ 3380 & 5.29 $\times 10^7  \pm$ 2.45 $\times 10^7$ & -2.07 $\times 10^7  \pm$ 2.14 $\times 10^7$ & -2.09 $\times 10^7  \pm$ 4.32 $\times 10^7$ & 2120 $\pm$ 7380 & 496 $\pm$ 5773 & 26.2 $\pm$ 18.6 & 173 $\pm$ 1.27 & 0.119 $\pm$ 0.001 & 0.424 $\pm$ 0.140 & 44.7 $\pm$ 11.0 & 58.2 $\pm$ 15.1\\
    \midrule
    MCMC: \verb|MCMC-diffusion| & -7160 $\pm$ 10 & 8.55 $\times 10^7  \pm$ 0.50 $\times 10^7$ & -9.91 $\times 10^7  \pm$ 0.08 $\times 10^7$ & 9.66 $\times 10^7  \pm$ 0.35 $\times 10^7$ & 1510 $\pm$ 990 & -9510 $\pm$ 490 & 69.9 $\pm$ 0.1 & 171 $\pm$ 0.01 & 0.120 $\pm$ 0.001 & 0.212 $\pm$ 0.013 & 63.0 $\pm$ 3.8 & 81.0 $\pm$ 3.74\\
    \midrule
    NS: \verb|Dynesty| & -7160 $\pm$ 10 & 8.57 $\times 10^7  \pm$ 0.49 $\times 10^7$ & -9.90 $\times 10^7  \pm$ 0.07 $\times 10^7$ & 9.66 $\times 10^7  \pm$ 0.34 $\times 10^7$ & 1500 $\pm$ 970 & -9520 $\pm$ 480 & 69.9 $\pm$ 0.1 & 171 $\pm$ 0.01 & 0.120 $\pm$ 0.001 & 0.213 $\pm$ 0.013 & 63.1 $\pm$ 3.8 & 81.3 $\pm$ 3.75\\
    \midrule
    NS: \verb|JaxNS| & -7160 $\pm$ 10 & 9.38 $\times 10^7  \pm$ 0.54 $\times 10^7$ & -6.45 $\times 10^6  \pm$ 5.45 $\times 10^6$ & 7.68 $\times 10^7  \pm$ 0.82 $\times 10^7$ & -9860 $\pm$ 130 & 9670 $\pm$ 310 & 42.4 $\pm$ 3.0 & 171 $\pm$ 0.10 & 0.120 $\pm$ 0.001 & 0.201 $\pm$ 0.001 & 66.5 $\pm$ 0.5 & 32.1 $\pm$ 1.08\\
    \midrule
    \rowcolor{row-highlight}
    NS: \verb|Nautilus| & -7162 $\pm$ 10 & 8.58 $\times 10^7  \pm$ 0.51 $\times 10^7$ & -9.91 $\times 10^7  \pm$ 0.07 $\times 10^7$ & 9.67 $\times 10^7  \pm$ 0.32 $\times 10^7$ & 1530 $\pm$ 1030 & -9490 $\pm$ 510 & 69.9 $\pm$ 0.1 & 171 $\pm$ 0.01 & 0.120 $\pm$ 0.000 & 0.213 $\pm$ 0.013 & 62.9 $\pm$ 3.9 & 81.1 $\pm$ 3.79\\
    \midrule
    NS: \verb|Ultranest| & -7160 $\pm$ 10 & 8.55 $\times 10^7  \pm$ 0.51 $\times 10^7$ & -9.91 $\times 10^7  \pm$ 0.07 $\times 10^7$ & 9.66 $\times 10^7  \pm$ 0.35 $\times 10^7$ & 1510 $\pm$ 1000 & -9510 $\pm$ 470 & 69.9 $\pm$ 0.1 & 171 $\pm$ 0.01 & 0.120 $\pm$ 0.001 & 0.213 $\pm$ 0.013 & 62.9 $\pm$ 3.9 & 81.2 $\pm$ 3.71\\
    \bottomrule
    \end{tabular}
    \caption{Posterior mean and error for the MSSM7 likelihood.}
    \label{tab:mssm7_means}
\end{table}
\end{landscape}
\newpage

\section{MCMC: Number of walkers}

\begin{table}[h]
    \hspace*{-1.5cm}
    \tiny
    \centering
    \begin{tabular}{ccccccc}
    \toprule

    Algorithm & Eggbox & Rosenbrock & Rastrigin & Spike-slab & Cosmology - Planck & Particle physics - LHC\\
    \midrule
    MCMC: \verb|zeus| & 100 & 50 & 30 & 30 & 12 & 12\\
    \midrule
    MCMC: \verb|ptemcee| & 100 & 100 & 30 & 30 & 12 & 12\\
    \midrule
    MCMC: \verb|emcee| & 100 & 2000 & 30 & 30 & 12 & 12\\
    \midrule
    MCMC: \verb|MCMC-diffusion| & 30 & 30 & 30 & 30 & 30 & 30\\

    \bottomrule
    \end{tabular}
    \caption{Number of MCMC walkers for each test problem.}
    \label{tab:walkers}
\end{table}

\bibliographystyle{JHEP}
\bibliography{HighDimensionalSampling}

\end{document}